\documentstyle[eqsecnum,aps]{revtex} 
\begin{document}

\thispagestyle{empty}
{\baselineskip-4pt
\font\yitp=cmmib10 scaled\magstep2
\font\elevenmib=cmmib10 scaled\magstep1  \skewchar\elevenmib='177
\leftline{\baselineskip20pt
\vbox to0pt{
}}

\rightline{\large\baselineskip20pt\rm\vbox to20pt{
\baselineskip14pt
\hbox{UAB-FT-418}
\hbox{OU-TAP 63}
\vspace{2mm}
\hbox{20 June 1997}
}}

\vspace{2cm} 

\begin{center}
{\Large\bf Canonical Quantization of Cosmological Perturbations}\\
{\Large\bf in the One Bubble Open Universe} 
\end{center}
\bigskip

\centerline{\large 
Jaume Garriga$^1$\footnote{Electronic address: 
garriga@ifae.es}, 
Xavier Montes$^1$\footnote{Electronic address: 
montes@ifae.es}, 
Misao Sasaki$^2$\footnote{Electronic address: 
misao@vega.ess.sci.osaka-u.ac.jp} 
and Takahiro Tanaka$^2$\footnote{Electronic address: 
tama@vega.ess.sci.osaka-u.ac.jp} }
\bigskip
\begin{center}
{\em $^1$IFAE, Edifici C,
  Universitat Aut{\`o}noma de Barcelona, 
E-08193 Bellaterra, Spain}\\
{\em $^2$Department of Earth and Space Science, 
Graduate School of Science,}\\ 
{\em  Osaka University, Toyonaka 560, Japan}\\
\end{center}

\bigskip

\begin{abstract}
  Faddeev and Jackiw's method for constrained systems is used to derive a
  gauge invariant formulation of cosmological perturbations in the one bubble
  inflationary universe. For scalar perturbations in a flat universe,
  reduction of the action to the one with a single physical degree of freedom
  has been derived in the literature.  A straightforward generalization of it
  to the case of an open universe is possible but it is not adequate for
  quantizing perturbations in the one bubble universe, because of the lack of
  Cauchy surfaces inside the bubble. Therefore we perform the reduction of the
  action outside the lightcone emanating from the center of the bubble or
  nucleation event, where the natural time constant hypersurfaces are no
  longer homogeneous and isotropic and as a result the conventional
  classification of perturbations in terms of scalar and tensor modes is not
  possible.  Nevertheless, after reduction of the action we find three
  decoupled actions for three independent degrees of freedom, one of which
  corresponds to the scalar mode and the other two to the tensor modes.
  Implications for the one bubble open inflationary models are briefly
  discussed. As an application of our formalism, the spectrum of long
  wavelength gravity waves is simply obtained in terms of the real part of the
  reflection amplitude for a one dimensional scattering problem, where the
  potential barrier is given in terms of the bubble profile.
\end{abstract}

\section{Introduction}

Models which reconcile inflation with a non-critical density, $\Omega\neq 1$,
have been recently proposed in the literature \cite{models}.  Although these
models are somewhat more involved than standard inflation, they may end up
being favored by observations \cite{ratra,topdef}.  In this scenario, one
starts with a scalar field trapped in a false vacuum that drives a de
Sitter-like phase of inflation.  This field undergoes a first order phase
transition by forming an O(3,1) symmetric bubble.  Inside the bubble, the
scalar field in the new phase slowly rolls down the potential, driving a short
second period of inflation. Our observable universe would be contained inside
a single bubble, whose symmetry accounts for the observed large scale
homogeneity and isotropy. The second period of inflation is needed to solve
the entropy problem.

A complete study of cosmological perturbations in open inflation involves the
quantization of fields in the presence of a bubble. So far, progress has been
made by quantizing the scalar field but ignoring the selfgravity of these
fluctuations. \cite{ratra2,sqfopen,cananis,beldle,cohn}. The quantization of
tensor modes has been considered in
\cite{allen,tamamilne,tamaopen,gwbuch,tamaanis,bell1}.  Some interesting
features have been found. There are some scalar modes - the so called
supercurvature modes- which are not normalizable in the open hyperboloids but
do contribute to the microwave background anisotropies. Some of these, which
correspond to fluctuations of the bubble wall, are found to be such that they
can be rewritten as tensor modes, due to their special eigenvalue of the
Laplacian \cite{k3,bell2,jaume}.  Of course, it would be very interesting to
know whether these features survive when gravitational perturbations are
included and all the perturbations are consider in a unified manner.

Gravitational perturbations contain gauge degrees of freedom.  Perhaps the
most elegant way to get rid of the ``unphysical'' gauge modes is the gauge
invariant theory of cosmological
perturbations.\cite{bardeen,kodsas,pertth,branden}.  In a universe dominated
by a single scalar field, there is only one physical degree of freedom for
scalar perturbations. Hence to quantize perturbations, one has to reduce the
number of variables in the action to a single variable by using the
constraints. This program has been carried out for scalar perturbations in a
spatially flat universe\cite{branden,qflat,mukhanov} and in a spatially closed
universe\cite{qclose}.  The final form of the action resembles the one for a
scalar field with a time-dependent mass term.  One might expect that the
extension of this program to the case of an open universe would be
straightforward. In fact, it would be so if the hypersurfaces of homogeneity
and isotropy of an open universe were Cauchy surfaces on which the canonical
commutation relations could be set up. However, in the case of one bubble
inflationary scenario, the whole universe is contained in a single bubble and
the open hypersurfaces foliate only the interior of the lightcone emanating
from the center of the bubble\cite{nocauchy}.  In particular, one cannot deal
with supercurvature modes on these hypersurfaces.  Hence the quantization
should be carried out outside the lightcone where Cauchy surfaces exist but no
hypersurface of homogeneity and isotropy exists.  The purpose of this paper is
to carry out this non-trivial task, i.e., to find the reduced action for
cosmological perturbations appropriate for the scenario of the one bubble
inflationary universe.

Dirac's procedure \cite{Dirac} has been for a long time the canonical way to
treat constrained systems. Faddeev and Jackiw (FJ), however, have
proposed\cite{tears} an alternative approach which leads to the same results
without following all of Dirac's steps. As they point out, two aspects of
Dirac's procedure can be avoided. First, it is not necessary to distinguish
between different classes of constraints: all of them can be treated on equal
footing without ambiguities.  Second, it is not necessary to define conjugate
momenta for those velocities which appear linearly in the Lagrangian, as is
customary done in Dirac's approach.  Applied to our case, the method gives the
linearized action for cosmological perturbations in terms of three gauge
invariant degrees of freedom, one corresponding to the scalar mode and two to
the tensor modes. This action is then ready for canonical quantization.

The paper is organized as follows. In Section II we describe the method of
reduction. In Section III we apply it to cosmological perturbations in open
inflation. In Section IV we derive, as an aplication of our formalism, the
spectrum of gravity waves in open inflation. In Section V we summarize our
conclusions. Some technical issues are left to the appendices.

\section{Reduction Method}

\label{sec:redmethod}

In the Faddeev-Jackiw approach one begins with an action first order in time
derivatives,
\begin{equation}
    S = \int\,\left(a_\mu(\xi, t)\dot\xi^\mu{} + L_0(\xi, z, t)\right)\,dt,
\end{equation}
where $\xi^\mu$ are the phase space variables of the system, and $z$ are a
subset of those which do not appear in the kinetic term. The basis of the
method\cite{tears} is to use the Euler-Lagrange equations of motion that
contain no time derivatives (the real constraints of the theory) to reduce the
phase space.  The equations of motion of the $z$ coordinates belong to this
category. One starts by solving this set of equations first for as many $z$'s
as possible and then, if there are any $z$'s that appear linearly in the
Lagrangian, for as many $\xi$'s as possible. After substituting these
relations into the original action, it takes the form
\begin{equation}
\label{eq:firstlag}
S^* = \int\,\left(f_i(\xi, t)\dot\xi^i + L^*_0(\xi, t)\right)\,dt,
\end{equation}
where now the label $i$ spans fewer coordinates than $\mu$. From now on, a
${}^*$ means that the known constraints have been substituted.

If there are further constraints in the theory, they manifest themselves as
combinations of the equations of motion which contain no time
derivatives. Writing the equations of motion from (\ref{eq:firstlag}) in the
form
\begin{equation}
\frac{\delta S^*}{\delta\xi^i} = f_{ij} \dot\xi^j{} + \frac{\partial
L^*_0}{\partial\xi^i} -\partial_t f_i =: f_{ij} \dot\xi^j{} + G_i(\xi,t)
=0,
\end{equation}
where $f_{ij} := \partial_i f_j - \partial_j f_i$, each zero mode $\xi^i_z$ of
the kinetic matrix $f_{ij}$, i.e. $\xi^i_z f_{ij} = 0$, will give us the
constraint equation
\begin{equation}
  \label{eq:constraint} \xi^i_z \,G_i(\xi,t) = 0.
\end{equation}
These constraints can be used again to reduce the phase space. The process is
repeated until we end up with a nonsingular $f_{ij}$, which indicates that we
have identified the reduced phase space in which the Lagrangian is
unconstrained. The equivalence of this method with Dirac's is discussed
e.g. in \cite{pons}.

{}For our present purpose, we do not have to follow the Faddeev-Jackiw
approach step by step.  In the problem of cosmological perturbations, the
constraints are first class and become the generators of gauge
transformations. In this specific case, we can make use of this fact for
reduction.  As is well known, the first order quadratic action for
gravitational perturbation takes the form,
\begin{equation}
   {\cal S} = \int\,\left(p_{\mu} \,\dot q^{\mu} 
       + p_a \dot q^a - {\cal H} - \delta{\cal N}^{\mu} \,
        C_{\mu}[p_{\mu},q^{\mu},Q]\right)\,d^4 x,
\end{equation}
where $C_{\mu}$ are the gauge transformation generators, i.e. the gauge
transformation of a variable is given by the Poisson bracket as $\delta_g f
=\biggl\{f,\int\,\lambda^{\mu}C_{\mu}\,d^3x\biggr\}$, 
and $Q^n= (q^a,p_a)$ stands for the rest of canonical coordinates. The
fields $\delta{\cal N}^{\mu}$ are the 
Lagrange multipliers corresponding to the perturbation of the lapse function
and the shift vector.  In our present problem, constraints $C_{\mu}=0$, which
are linear in the perturbed variables, are easily solved for $p_{\mu}$ as
$p_{\mu}=\hat p_{\mu}[q^{\nu},Q^n]$.  That is, by taking a linear
combination of $C_{\mu}$, we can rewrite the constraints as $\tilde C_{\mu} =
p_{\mu}-\hat p_{\mu}[q^{\nu},Q^n]$.  Accordingly, we write $\delta_g f
=\biggl\{f,\int \,\tilde \lambda^{\mu} \, \tilde C_{\mu}\,d^3x \biggr\}$.
Substituting
$p_{\mu}=\hat p_{\mu}[q^{\nu},Q^n]$ back into the action, we obtain
\begin{equation}
   {\cal S}^* = \int \,\left(\hat p_{\mu}[q^{\nu},Q^n] \,\dot q^{\mu} 
         + p_a \dot q^a - {\cal H}^*\right)\,d^4 x, 
  \hskip 1cm {\cal H}^* = {\cal H}|_{p_{\mu}=\hat p_{\mu}[q^{\nu},Q^n]},
\end{equation}
which loses its canonical form but still keeps gauge invariance.  
This can be deduced from the well known fact that the constraints 
$\tilde C_{\mu}$ are first class: 
$\{\tilde C_{\mu},\tilde C_{\nu}\}=0$.\footnote{In general, 
this equality holds only in a weak sense. 
But, in our present problem, $\tilde C_{\mu}$ are linear in 
the variables and hence their Poisson brackets are just numbers. 
Therefore this weak equality can be interpreted as a strong one.} 
Then we find 
\begin{equation}
 \delta_g p_{\mu}= \left\{\hat p_{\mu}[q^{\nu},Q^n], 
   \int \tilde\lambda^{\alpha}\tilde C_{\alpha}\, d^3 x \right\}.
\end{equation}  
Now, we can show the gauge invariance of ${\cal S}^{*}$ as 
\begin{eqnarray}
  \delta_g {\cal S}^{*}&=&\left\{\hat p_{\mu}[q^{\nu},Q^n], \int \tilde
    \lambda^{\alpha} \tilde C_{\alpha}\, d^3 x \right\} 
    \left.{\delta {\cal S}\over \delta p_{\mu}}\right|_{\tilde C_\nu=0} 
  + \delta_g q^\mu\,
  \left.\frac{\delta {\cal S}}{\delta q^\mu}\right|_{\tilde C_\nu=0}
 + \delta_g Q^n
  \left.{\delta {\cal S}\over \delta Q^n}\right|_{\tilde C_\nu=0} \nonumber
  \\ &=&\delta_g {\cal S}\bigg\vert_{\tilde C_\nu=0}=0.
\end{eqnarray}

Writing the gauge transformation of a given variable in terms of operators
acting on the gauge parameters, i.e. $\delta_g Q^n =:
\widehat{\delta Q^n_\mu} [\tilde \lambda^\mu]$, gauge invariance can
also be written as
\begin{equation}
  \delta_g{\cal S}^* = \int\,d^4 x\,\left({\tilde
      \lambda}^\mu\,\frac{\delta}{\delta q^\mu(x)} + \widehat{\delta
      Q^n_\mu}[\tilde\lambda^\mu]\frac{\delta}{\delta Q^n
      (x)}\right){\cal S}^* = 0,
\end{equation}
which implies that
\begin{equation}
  \label{eq:idfield} \frac{\delta {\cal S}^*}{\delta q^\mu} = -
 \widehat{\delta Q^n_\mu}^\dagger[\frac{\delta{\cal S}^*}
{\delta Q^n}],
\end{equation}
where $\widehat{\delta Q^n_\mu}^\dagger$ is the operator conjugate to
$\widehat{\delta Q^n_\mu}$ defined as
\begin{equation}
  \label{dagdef}
\int d^4x \widehat{\delta Q^n_\mu}^\dagger[f]\,g
:=\int d^4x\,f\,\widehat{\delta Q^n_\mu}[g],
\end{equation}
for any square-integrable functions $f$ and $g$.
Thus, recalling the chain rule, ${\cal S}^{*}$ should depend on $q^\mu$
only through the following combination of variables
\begin{equation}
  {\bbox{Q}}^n := Q^n - \widehat{\delta Q_\mu^n}[q^\mu]
  = Q^n - \left\{Q^n, \int q^{\mu} \tilde C_{\mu}\,d^3
    x\right\}.
\label{giv}
\end{equation}
It is easily shown that the new set of reduced variables ${\bbox{Q}}^n$ is
gauge invariant.  In an actual calculation, all we have to do is just to set
$q^{\mu}=0$ in ${\cal S}^{*}$ and reinterpret $Q^n$ as their gauge
invariant counterparts, ${\bbox{Q}}^n$.  In this way, we go from the set
of variables $(Q^{n},q^{\mu},p_{\mu})$ to the reduced set ${\bbox{Q}}^n$.

Equation (\ref{eq:idfield}) can be also derived directly from the full
Faddeev-Jackiw procedure. If instead of taking advantage of the gauge
invariance we followed the method step by step, after
substituting the four constraints $\tilde C_\mu$ we would find that the
kinetic matrix has four zero modes. The constraint equations
(\ref{eq:constraint}) for these zero modes turn out to be
\begin{equation}
   \frac{\delta {\cal S}^*}{\delta q^\mu} +
 \widehat{\delta Q^n_\mu}^\dagger[\frac{\delta{\cal S}^*}
{\delta Q^n}] = 0,
\end{equation}
but, as we have seen, due to the gauge invariance, the left-hand side
vanishes identically. As explained, these identities point out that the
action can be written in terms of the reduced gauge invariant set of
variables (\ref{giv}).

\section{Open Inflation}

As mentioned in the introduction, the interior of an O(3,1) symmetric bubble
is isometric to an open Friedmann-Robertson-Walker (FRW)
universe. Unfortunately, the hypersurfaces of homogeneity and isotropy of this
universe are not appropriate for setting canonical commutation relations or
normalizing modes \cite{nocauchy}, because they are not Cauchy surfaces for
the whole space time.  Therefore, we shall need to quantize on spacelike
hypersurfaces which cut right through the bubble, and which are therefore not
homogeneous. This renders the decomposition into scalar, vector and tensor
modes into a somewhat unfamiliar form. In the end, however, the three standard
physical degrees of freedom will be identified.

The open FRW chart
\begin{equation}
  ds^2 = -dt^2 + a(t)^2 d\Omega_{H_3}\,,
 \hskip 2cm d\Omega_{H_3}=dr^2 + \sinh^2
  r(d\theta^2 + \sin^2 d\varphi^2),
\end{equation}
covers only the interior of the lightcone emanating from the center of
the bubble $t=0$, $r=0$, which we shall call the nucleation event N.
Following \cite{sqfopen}, we shall call the interior of the lightcone
region R and the outside of this lightcone region C. Region C can be
covered by analytically continuing the coordinates $t$ and $r$ in region 
R to the complex plane. By taking $t=i\tau$ and $r=\chi + i(\pi/2)$,
with $\tau$ and $\chi$ real, the line element becomes
\begin{equation}
  ds^2 = d\tau^2 + a_{E} (\tau)^2 d\Omega_{dS}, \hskip 1cm
  d\Omega_{dS}=\gamma_{ij}^{dS}dx^i dx^j = -d\chi^2 + \cosh^2 \chi(d\theta^2 +
  \sin^2 d\varphi^2),
\end{equation}
where $a_{E}(\tau)=-i\,a(i\,\tau)$, and $d\Omega_{dS}$ is the metric of a
(2+1) dimensional de Sitter space. In this chart, $\tau$ is a `radial'
spacelike coordinate, whereas $\chi$ is timelike. Now the spacelike
hypersurface $\chi = 0$ is a Cauchy surface for the entire
space-time\cite{nocauchy}.  It is convenient to introduce the conformal
`radial' coordinate ${\eta_{E}}$, with $d\tau = -a_{E}\,d{\eta_{E}}$. Close to
the lightcone emanating from the nucleation event N, the scale factor behaves
as $a_{E}(\tau)\approx \tau$, and $\eta_{E}\to +\infty$. As we move away from
N along the $\tau$ direction, the scale factor rises to a maximum and then
decreases again, reaching another zero at the so called antipodal point A,
which corresponds to $\eta_{E}\to -\infty$.

Writing the perturbed line element and the perturbed scalar field in the form
\begin{equation}
\label{eq:linepertout}
\begin{array}{lcl}
  ds^2 & = & a_{E}({\eta_{E}})^2\{(1 + 2 A)d\eta_{E}^2 - 2 S_{i}dx^i
  d{\eta_{E}} +
  (\gamma_{ij}^{dS} - h_{ij})dx^i dx^j\} \nonumber,\\
  \varphi & = & \varphi_0({\eta_{E}}) + \delta\varphi,
\end{array}
\end{equation}
the second order action for small perturbations is given by
\begin{eqnarray}
  \label{eq:totalactioncont} \delta_2{\cal S} &=& - \frac{1}{2 \kappa} \int
  d^4x \,a_{E}^2 \sqrt{-\gamma^{dS}}\{-2(2 {\cal H}^2 + {\cal H}') A^2 +
  (S_{(i|j)} - \frac{h'_{ij}}{2})^2 - (S^i{}_{|i} -
  \frac{h^i{}_{i}{}'}{2})^2\nonumber \\ & & \hskip 1 cm - \frac{1}{4}(2
  h^{ij|k}h_{jk|i} - h^{ij|k}h_{ij|k} - 2 h^{ij}{}_{|j} h^k{}_{k|i} +
  h^i{}_i{}^{|j}h^k{}_{k|j}) +  A^{|i}(h^j{}_{j|i} - h^j{}_{i|j})
  \nonumber\\ & & \hskip 1 cm +  \kappa (\delta\varphi'^2 +
  \delta\varphi^{|i}\delta\varphi_{|i} + a_{E}^2 V_{,\varphi\varphi}
  \delta\varphi^2) - 2 \kappa(\varphi_0' \delta\varphi' A - a_{E}^2
  V_{,\varphi} \delta\varphi A)\nonumber\\ & & \hskip 1 cm - 4 (S^i{}_{|i} -
  \frac{h^i{}_{i}{}'}{2})(\frac{\kappa}{2} \varphi_0' \delta\varphi - 
  {\cal H} A) - (2
   A^2 + 2  A h^i{}_i - \frac{1}{2} h^i{}_i h^j{}_j + h^{ij} h_{ij})\},
\end{eqnarray}
where $\kappa=8\pi\,G$, ${\cal H} = a'/a$ and a prime denotes a derivative
with respect to the conformal `radial' coordinate $\eta_{E}$. This can be found
e.g. from Appendix~\ref{sec:seconorderin} just replacing $a^2\rightarrow
-a_{E}^2$ and $\gamma_{ij} \rightarrow - \gamma_{ij}^{dS}$ in expression
(\ref{eq:totalaction}).

As mentioned before, the usual expansion of the metric perturbations (see
e.g. \ref{eq:decomposition}) in scalar, vector and tensor modes with respect
to the 3-hyperboloid, used inside the lightcone, cannot be used outside. 
The reason is that, in region C, the corresponding 3-hyperboloid on which
these harmonics of various types are defined no longer gives a spatial section
of the spacetime.  Instead we shall expand in scalar and vector modes with
respect to the 2-sphere. Using a conformal time-like coordinate $\rho$ defined
through the relation $\cosh \chi \,d\rho = d\chi$, the metric element for the
de Sitter space can be written as
\begin{equation}
  ds^2_{dS} = c_{E}(\rho)^2(-d\rho^2 + \omega_{AB}dx^A dx^B),
\end{equation}
with $c_{E}(\rho) \equiv \csc \rho$, and where $A$, $B$, ...
 run over $\theta$ and $\varphi$. For convenience, we define 
$h_{E} = \dot c_{E}/c_{E}$, where a dot
indicates derivative with respect to the time-like variable $\rho$.

We write the metric perturbations as
\begin{eqnarray}
  \label{eq:expout} S_\rho &=& - S \nonumber, \\
  S_A &=& T_{||A} + V_A\nonumber,\\
  h_{\rho\rho} &=& 2 \,c_{E}^2 \zeta, \\
  h_{\rho A} &=& - c_{E}^2\,(\xi_{||A} + W_{A}) \nonumber,\\
  h_{AB} &=& c_{E}^2 \,((w + {^{(2)}\triangle}v)\omega_{AB} - 
    2 \,v_{||AB} + 2 F_{(A||B)})\nonumber,
\end{eqnarray}
where $_{||A}$ and ${^{(2)}\triangle}$ stand for the 
covariant derivative and scalar Laplacian associated with $\omega_{AB}$, 
respectively. The meaning of ${^{(2)}\triangle}$ when operates on 
a vector or tensor quantity is explained immediately below. 
The fields $S$, $T$, $\zeta$, $\xi$, $w$ and $v$ are scalar modes,
and $V_A$, $W_A$ and $F_A$ are divergenceless vector modes with respect 
to the 2-sphere metric $\omega_{AB}$. More explicitly, we can write, say, 
$S=\sum_{lm}S^{\ell m}(\rho)Y^{\ell m}(\Omega)$ for scalar modes
and $V_A=\sum_{\ell m} V^{\ell m}(\rho)\epsilon^B_{~A} 
Y^{\ell m}_{||B}(\Omega)$ for vector modes, 
 where $Y^{\ell m}(\Omega)$ are the ordinary spherical harmonics which 
satisfy $^{(2)}\triangle Y^{\ell m}(\Omega)=-l(l+1) 
 Y^{\ell m}(\Omega)$, $\epsilon_{AB}$ is the unit anti-symmetric tensor
 on the unit 2-sphere ($\epsilon_{\theta\varphi}=\sin\theta$ etc.) and 
$\epsilon^B_{~A}=\omega^{BC}\epsilon_{CA}$. 
Now we can clearly state the meaning of $^{(2)}\triangle$. 
It should be understood just as $-l(l+1)$ when it is decomposed 
into modes. In order to avoid writing the 
summation over $l$ and $m$ for notational simplicity, we use 
$^{(2)}\triangle$ instead of $-l(l+1)$.
  
The scalar and vector modes transform differently under the 
parity transformation. 
Thus, inserting the decomposition
(\ref{eq:expout}) into the action (\ref{eq:totalactioncont}), 
the scalar and vector modes decouple, so they evolve independently.
According to the change of signature under the parity transformation, 
we refer to the scalar (vector) modes as even (odd) parity modes. 

We shall see that the odd parity modes contain one
physical degree of freedom, which corresponds to odd parity tensor modes
when analytically continued inside the lightcone
\cite{tamamilne,tamaopen}.
  The even parity modes contain two degrees of freedom,
one corresponding to the usual scalar and the other to even parity tensor
modes.

\subsection{Odd parity modes}
\label{sec:oddparity}

First we consider the odd parity modes.  The result provided in this
subsection is essentially the same as that given in \cite{tamaopen}.  However,
our present approach based on the FJ method is quite different from the
conventional Dirac's method used in the previous work \cite{tamaopen}.
Furthermore, the analysis of the even parity modes discussed in the next
subsection is rather complicated compared with the odd parity modes.  So, also
to understand our strategy, it will be convenient to present the analysis of
the odd parity modes first.

The Lagrangian density for odd parity modes is
\begin{eqnarray}
  \label{actvect} 
^{(v)}{\cal L} &=&\frac{a_{E}^2\sqrt{\omega}}{4c_{E}\kappa}
 \left\{(\dot V_A - 2h_{E} V_A+ c_{E}^2 W_A')^2 
+(V^A - c_{E}^2 F^A{}')({^{(2)}\triangle}+2)(V_A - c_{E}^2F_A') \right.
 \nonumber\\ & & \left.- c_{E}^2 (\dot F^A + W^A)
 ({^{(2)}\triangle}+2)(\dot F_A + W_A)\right\}.
\end{eqnarray}
By the definition of conjugate momenta, we find  
\begin{equation}
\label{oddmom}
\begin{array}{lcl}
  \Pi_V^A &:=&
 \displaystyle \frac{\partial\,{}^{(v)}{\cal L}}{\partial \dot V_A} 
= \frac{a_{E}^2 \sqrt{\omega}}{2\kappa c_{E}}(\dot V^A- 2 h_{E} V^A +
  c_{E}^2 W^A{}'), \\
  \Pi_F^A &:=& 
\displaystyle \frac{\partial\,{}^{(v)}{\cal L}}{\partial \dot F_A}
=  -\frac{a_{E}^2c_{E}\sqrt{\omega}({^{(2)}\triangle}+2)}{2\kappa}
    (\dot F^A + W^A).
\end{array}
\end{equation}
Here we have raised indices with $\omega^{AB}$. 
If we cast it into first order form we obtain
\begin{eqnarray}
  \label{actvectfirst} 
{}^{(v)}{\cal L} &=& {}^{(v)}{\cal L}_1 - {}^{(v)}{\cal H}
  - {\cal C}_W^A W_A, \\
  ^{(v)}{\cal L}_1 
  &=& \Pi_V^A \dot V_A + \Pi_F^A \dot  F_A\nonumber \nonumber,\\ 
  {}^{(v)}{\cal H} &=&
  -\Pi_F^A\frac{\kappa}{a_{E}^2c_{E}\sqrt{\omega}({^{(2)}\triangle}+2)} 
\Pi_F{}_A + \frac{\kappa c_{E}}{a_{E}^2\sqrt{\omega}}\,\Pi_V^A\Pi_V{}_A + 
  2h_{E} \Pi_V^A V_A\nonumber\\ 
  & &- \frac{a_{E}^2\sqrt{\omega}}{4\kappa c_{E}}(V^A- c_{E}^2
  F^A{}')({^{(2)}\triangle} +2)(V_A- c_{E}^2 F_A')  \nonumber,\\
  {\cal C}_W^A &=& - \Pi_F^A + c_{E}^2 \Pi_V^A{}'. \nonumber
\end{eqnarray}

To find the reduced phase space of the lagrangian (\ref{actvectfirst}), we
solve ${\cal C}_W^A = 0$ for $\Pi_{F}^A$, and substitute it back into the
Lagrangian.  Using the fact that ${\cal C}_W$ is the generator of odd
parity gauge transformations (see Appendix \ref{sec:gauoutside}), the
prescription given in (\ref{giv}) indicates that the gauge invariant
combinations are given by ${\bbox{V}}_A:=a_{E}^2\sqrt{\omega}\,(V_A -
c_{E}^2 F'_A)$ and ${\bf \Pi}^A:=\Pi_{V}^A/(a_{E}^2\sqrt{\omega})$. 
The canonical first order Lagrangian for this degree of freedom is:
\begin{eqnarray}
  {}^{(v)}{\cal L}^* &=& {\bf \Pi}^A \dot{\bbox{V}}_A + \kappa \,c_{E}a_{E}^2
  \sqrt{\omega} \left( c_{E}^2 {\bf \Pi}^A{}' { 1\over ({^{(2)}\triangle} +
      2)} {\bf \Pi}'_A - {\bf \Pi}^A {({^{(2)}\triangle}+2 + 2 c_{E}^2 {\cal
        H}') \over ({^{(2)}\triangle} + 2)} \, {\bf \Pi}_A\right)
  \nonumber \\
  & & -\,2h_{E} {\bf \Pi}^A {\bbox{V}}_A + \frac{{\bbox{V}}^A
    ({^{(2)}\triangle} + 2){\bbox{V}}_A}{4 \kappa \,c_{E} a_{E}^2
    \sqrt{\omega}}.
\end{eqnarray}
Solving for the velocity $\dot {\bbox{V}}_A$, we find the second order reduced
action for ${\bf \Pi}_A$. It is convenient to express the divergenceless
vector as ${\bf \Pi}_A=:{\bf \Pi}^{||B}\epsilon_{BA}$.  With this we
obtain
\begin{eqnarray}
  {}^{(v)}{\cal S}^{(2)} 
&=& 2\kappa\int\frac{\sqrt{-^{(4)}g}}{2}
          \,{\bf \Pi}\left\{\frac{{^{(2)}\triangle}}
          {{^{(2)}\triangle} + 2} \left(\Box + \frac{
          2{\cal H}'}{a_{E}^2}\right)\right\}{\bf \Pi}\,d^4x
\nonumber\\
&=&2\kappa\int\frac{\sqrt{-^{dS}\gamma}}{2}
          \,(a_{E}{\bf \Pi})\left\{\frac{{^{(2)}\triangle}}
       {{^{(2)}\triangle} + 2} \left(^{dS}\Box -1-{\widehat K}\right)\right\}
       (a_{E}{\bf \Pi})\,d\eta_Ed^3x,
 \label{bigone}
\end{eqnarray}
where $\Box$ stands for the four dimensional d'Alembertian,
$^{dS}\Box$ is the d'Alembertian on the (2+1) dimensional
de Sitter space, and ${\widehat K}$ is the operator defined as
\begin{equation}
  {\widehat K}:= -\frac{d^2}{d{\eta_{E}}{}^2} +\frac{\kappa}{2} \varphi_0'{}^2.
\label{Optensor}
\end{equation}
The above action is
really very simple when expanded in eigenmodes. If one absorbs the
factor 
\begin{equation}
{2\kappa\,{}^{(2)}\triangle\over(^{(2)}\triangle+2)}=
 {2\kappa\,{}\ell(\ell+1)\over (\ell+2)(\ell-1)}=:
{1\over ({\cal N}_{-}^{\ell})^{2}},
\end{equation}
by redefinition of
${\bf \Pi}$, we basically obtain the action for 
an ordinary scalar field, $\check {\bbox{\Pi}}$, 
living in the curved background describing the bubble geometry, with an
$\eta_E$-dependent mass term. 

We decompose the field $\bf \Pi$ into modes as
\begin{equation}
  \label{Pidecomp}
  {\bf \Pi}=\sum_{p\ell m}{\cal N}_{-}^{\ell}
    \bbox{\pi}_{p\ell m} U_T^{p\ell m}(x),
\end{equation}
where $\bbox{\pi}_{p\ell m}$ is the coefficient 
which represents the amplitude and 
$U_T^{p\ell m}(x)$ is a suitably normalized mode function 
which is also an eigenfunction 
of the operators ${\widehat K}$ and $^{(2)}\triangle$. 
Then the renormalized field $\check {\bbox{\Pi}}$ 
is defined by $\check {\bbox{\Pi}}:=
\sum_{p\ell m}\bbox{\pi}_{p\ell m} U_T^{p\ell m}(x)$. 
It is convenient to normalize $U_T^{p\ell m}(x)$
by means of the Klein-Gordon norm with respect to 
the renormalized field $\check{\bbox{\Pi}}$. 
With this choice of normalization, when we 
go to the quantum theory by setting the canonical 
commutation relation between the operator counter part of 
$\bbox{\Pi}$ and its conjugate, $\bbox{\pi}_{p\ell m}$ 
can be recognized as the anihilation operator which 
satisfies $[\bbox{\pi}_{p\ell m},\bbox{\pi}_{p\ell m}^{\dag}]=1$. 
Writting $U_T^{p\ell m}$ in the form,
\begin{equation}
\label{UTdef}
  U_T^{p\ell m}(x)= a_{E}^{-1}
 {u}^{p}({\eta_{E}}){\cal Y}^{p\ell m}(x^i),
\end{equation}
 the equation of motion separates into
\begin{eqnarray}
  ^{dS}\Box{\cal Y}^{p\ell m} &=& (p^2+1) \,{\cal Y}^{p\ell m},
  \label{calYeq}\\ 
  {\widehat K}[{u}^{p}]&=& p^2\,{u}^{p}.
\label{upspec}
\end{eqnarray}
These equations 
admit the following interpretation. The first tells us
that ${\cal Y}^{p\ell m}$ behave as scalar fields of mass $(p^2+1)$
living in a (2+1) dimensional de Sitter spacetime.  The spectrum of
masses is determined by (\ref{upspec}), 
which is a one dimensional Sch{\"o}dinger equation with effective
potential ${\cal U}_{\rm eff} = \kappa \varphi_0'{}^2/2$.  Notice that the
potential in (\ref{upspec}) is positive definite and vanishes at
infinity. Hence, the spectrum is purely continuous, with
$p^2>0$. Therefore, our formalism bypasses the issue of a possible
supercurvature mode discussed in \cite{tamaopen}. That mode was shown to
be pure gauge, so it is not surprising that it does not 
arise with a suitable choice of variable which contains no gauge degree of
freedom. 
If we choose the modes ${\cal Y}^{p\ell m}$ to be Klein-Gordon normalized
in the (2+1) dimensional sense, i.e.,
\begin{equation}
  \label{calYnorm}
  i\int d^2x\sqrt{\omega}\,c_{E}
 \left(\dot{\cal Y}^{p\ell m}\overline{{\cal Y}^{p\ell'm'}}
      -{\cal Y}^{p\ell m}\overline{\dot{\cal Y}^{p\ell'm'}}\right)
=\delta_{\ell\ell'}\delta_{mm'}\,,
\end{equation}
then 
the (3+1) normalization condition reduces to
\begin{equation}
  \label{upnorm}
\int^{\infty}_{-\infty}{u}^{p}\overline{u^{p'}}d\eta_E
 = \delta(p-p'),
\end{equation}
which is the standard one for the Schr{\"o}dinger problem.
For definiteness, here we choose 
\begin{equation}
 {\cal Y}^{p\ell m}:={i^{\ell+1}\Gamma(ip+\ell+1)\over\sqrt{2}}
   {\cal P}_{p\ell}(\rho) Y^{\ell m}(\Omega), 
\label{calYdef}
\end{equation}
where ${\cal P}_{p\ell}$ is defined by using the associated Legendre 
function of the first kind as 
\begin{equation}
 {\cal P}_{p\ell}(\rho):={P_{ip-{1\over 2}}^{-\ell-{1\over 2}}(i h_{E})\over 
     \sqrt{i c_{E}}}. 
\end{equation}

Notice that the factor ${^{(2)}\triangle} + 2$ becomes zero when $\ell =1$.
In this case, we have to go back to the original Lagrangian
(\ref{actvect}). From Eq.~(\ref{oddmom}) we find $\Pi^p_F=0$, which
means that one extra constraint arises.  Therefore there remain no
physical degrees of freedom for $\ell=1$ mode.  In fact, this case can
be quickly treated along the lines of Faddeev-Jackiw
approach. Substituting ${^{(2)}\triangle} + 2=0$ into the Lagrangian
(\ref{actvect}), and casting it into first order form, we obtain
\begin{eqnarray}
  \label{leqone}
  {}^{(v)}{\cal L}_{\ell=1} = \Pi_V^A\dot V_A +
  \frac{\kappa c_{E}}{a_{E}^2\sqrt{\omega}} \Pi_V^A\Pi_{V}{}_A
 -  2h_{E}\,V_A \Pi_V^A - c_{E}^2\,W_A\Pi_V^A{}' ,
\end{eqnarray}
which is independent of $F_A$.  The variation with respect to $W_A$
gives the constraint $\Pi_V^A{}'=0$. Then the normalizability of the
mode functions requires $\Pi_V^A=0$. 
 After substituting this constraint, the equation of
motion for $V_A$ becomes also a constraint, which enforces the Lagrangian for
$\ell=1$ to vanish. Modes with $\ell=0$ are also absent from the action by
construction. The absence of modes with $\ell=0,1$ is what we expect, because
the odd parity mode represents one of the tensor degrees of freedom inside the
lightcone, for which the modes $\ell=0,1$ do not exist.

Now we relate the quantities in the outside of the lightcone 
with those in the inside of it.  
Inside the lightcone 
we can use the tensor harmonics to decompose 
the tensor part of the metric perturbation into modes. 
Thus the mode function $U^{(-)}_{p\ell m}$ defined 
in Appendix B\cite{tamaopen} will be the most convenient choice 
of the variable to specify the tensor perturbation there. 
In order to relate the amplitude $\bbox{\pi}_{p\ell m}$ to 
$U^{(-)}_{p\ell m}$, we compare the $(\rho A)$-component 
of the metric perturbation in the synchronous gauge 
$(V_p=0)$. Following the notation in Appendix D, we associate 
a subscript (or superscript) $N$ to indicate the quantity 
evaluated in this gauge. 
{}From Eq.~(\ref{oddmom}), 
$h_{\rho A}^{N}$ is evaluated as 
\begin{eqnarray}
  {h^N}'_{\rho A} &=& -2\kappa c_{E} {\bf \Pi}_A \cr
  &=& \sum_{p\ell m} \displaystyle 
\frac{2\kappa c_{E}{\cal N}_{-}^{\ell}\bbox{\pi}_{p\ell m}}
    {a_{E}}\epsilon_{A}{}^{B} {\cal Y}^{p\ell m}_{||B} u^p\,
\end{eqnarray}
On the other hand, 
the explicit expression for $(\rho A)$-component of the 
tensor harmonics is given in \cite{tomita,tamaopen}. 
After the analytic continuation to region C, 
for the odd parity 
graviatational wave perturbation, we obtain 
\begin{equation}
 h_{\rho A}^{N}=\sum_{p\ell m} 
   \sqrt{(\ell-1)(\ell+2)\Gamma(ip+\ell+1)
   \Gamma(-ip+\ell+1)\over 2p^2(p^2+1)\Gamma(ip)\Gamma(-ip)} c_{E} 
   {\cal P}_{p\ell} \epsilon_A{}^{B} Y^{\ell m}_{||B} U^{(-)}_{p\ell m}. 
\end{equation}

Hence we find that the amplitude 
$\bbox{\pi}_{p\ell m}$ is related to the 
variable $U^{(-)}_{p\ell m}$ by
\begin{equation}
  \bbox{\pi}_{p\ell m}{u}^{p} =
  \frac{1}{\sqrt{2\kappa}\,i^{\ell+1}}\sqrt{\frac{\Gamma(-ip+\ell+1)}
{(p^2+1)\Gamma(ip+\ell+1)\Gamma(ip)\Gamma(-ip)}}\,a_{E} 
{d U^{(-)}_{p\ell m}\over d\eta_{E}}\,.
\end{equation}
Conversely, by using the equation satisfied by $U^{(-)}$ of
ref\cite{tamaopen},
\begin{equation}
\left({1\over a_{E}^2}{d\over d\eta_{E}}
a_{E}^2{d\over d\eta_{E}}+(p^2+1)\right)U^{(-)}=0,
\label{Ueq}
\end{equation} 
$U^{(-)}$ is expressed in terms of $\bbox{\pi}$ as
\begin{equation}
  U^{(-)}_{p\ell m}=-\sqrt{2\kappa}\,i^{\ell+1}\,
\sqrt{\frac{\Gamma(ip+\ell+1)\Gamma(ip)\Gamma(-ip)}
{(p^2+1)\Gamma(-ip+\ell+1)}}\,a_{E}^{-2}\bbox{\pi}_{p\ell m}
{d(a_{E}{u}^{p})\over d\eta_{E}}\,.
\end{equation}
 
\subsection{Even parity modes}

The even parity modes contain two dynamical degrees of freedom. One of
them is the usual scalar mode, and the other is the even parity tensor mode
discussed in \cite{tamaopen}.  After lengthy algebra, complicated by the
fact that the spatial sections are not homogeneous outside the lightcone
from the nucleation event, the Lagrangian can be cast into second order
form as the sum of a Lagrangian for the scalar mode plus a Lagrangian
for the even parity tensor mode.  The details of the reduction of the
action are given in Appendix C. 
Here we only discuss the meaning of the final results.

For the scalar part, we have
\begin{equation}
  \label{eq:mainoutsc} 
 {\cal S}^{(2)}_{\bbox{q}} =
  \frac{1}{2}\int\sqrt{-\gamma^{dS}}\, ({\widehat{\cal O}}\,{\bbox{q}} ) 
  \left\{^{dS}\Box+3-{\widehat{\cal O}}\right\}{\bbox{q}} \, d\eta_{E}\,d^3x,
\end{equation}
where we have introduced the Schr{\"o}dinger-like operator
\begin{equation}
  {\widehat{\cal O}}:= - \frac{d^2}{d\eta_{E}^2} + \frac{\kappa}{2}
  \varphi_0'{}^2 + \varphi_0'\left(\frac{1}{\varphi_0'}\right)'',
  \label{os}
\end{equation}
where $^{dS}\Box$ stands as before for the
d'Alembertian on the (2+1) dimensional de Sitter space of unit radius. The
variable ${\bbox{q}} $ is related to the gauge invariant potential
${\bf \Phi}_{\bf H}$ of Bardeen\cite{bardeen} 
when evolved to the outside of the lightcone 
(see Appendix~\ref{purely}):\footnote{The potential ${\bf \Phi}_{\bf H}$
  is given by ${\bf \Phi}_{\bf H} :=- \psi + (B-E')$ (see 
Appendix~\ref{sec:redins}).
 We also recall that ${\bf \Phi}_{\bf H}$ is equal to ${\bf \Phi}$ of
 Kodama and Sasaki\cite{kodsas} and to $-\bf \Psi^{(gi)}$ of
Mukhanov, Feldmann and Brandenberger\cite{branden}.}
\begin{equation}
  \label{eq:qpsi}
  {\bbox{q}}  = - \frac{2\,a_{E}}{\kappa\varphi_0'}\,{\bf \Phi}_{\bf H}.
\end{equation}

Putting ${\bbox{q}}=a_E 
\sum{\cal N}^p{\bbox{q}}_{p\ell m} U_S^{p\ell m}(x)$ 
with the mode function of the form $U_S^{p\ell m}
 = a_E^{-1}q^p({\eta_{E}})\,{\cal Y}^{p\ell m}(x^i)$, 
the equation of motion separates into (\ref{calYeq}) and 
\begin{eqnarray}
 {\widehat{\cal O}}[q^p]= (p^2 + 4)\,q^p.
\label{eq:schlike}
\end{eqnarray}
Just as in subsection \ref{sec:oddparity}, 
the masses $(p^2+1)$ of the (2+1)
dimensional fields ${\cal Y}^{p\ell m}$ are determined as the
eigenvalues of the Schr{\"o}dinger equation (\ref{eq:schlike}).
Now, if we absorb the factor 
\begin{equation}
{\widehat{\cal O}}=p^2+4=:{1\over ({\cal N}^p)^{2}}, 
\end{equation}
by defining 
$\check{\bbox{q}}:=\sum \bbox{q}_{p\ell m} U_S^{p\ell m}$, 
we obtain the action for an ordinary scalar field. 
As before, we require that $U_S^{p\ell m}$ is normalized 
with respect to the Klein-Gordon norm for the renormalized 
field $\check{\bbox{q}}$. This normalization condition  
reduces to
\begin{equation}
  \int^{\infty}_{-\infty}q^p\overline{q^{p'}}d\eta_{E}
 = \left\{\begin{array}{ll}
\delta(p-p')\quad&\hbox{for continuous $p$},\\
\delta_{pp'}\quad&\hbox{for discrete $p$}.\\
\end{array}\right.
\end{equation}
Since the potential of the operator ${\widehat{\cal O}}$ is not positive
definite, we cannot determine the spectrum of $p^2$ unless we solve
Eq.~(\ref{eq:schlike}). If the spectrum obtained previously by ignoring
degrees of freedom of the metric perturbations \cite{sqfopen} does not change
(except for the wall fluctuation mode at $p^2=-4$; see below), the spectrum
will be continuous for $p^2>0$ and there may be one discrete mode at
$-1<p^2<0$. We will discuss this issue in a forthcoming paper\cite{stspec}.

Note that for the modes with $p^2=-4$ we have ${\widehat{\cal O}}[q]=0$. If
one ignores the metric perturbations, these correspond to the wall fluctuation
modes \cite{sqfopen,k3,bell2,jaume}. Once the metric perturbations are taken
into account, however, the wall fluctuation modes are found to be contained in
the continuous spectrum of the gravitational wave perturbations
\cite{tamaanis}.  Hence we expect the modes with $p^2=-4$ cease to contribute
to physical fluctuations. In fact, there is a strong evidence that this is
true by examining the regularity of the metric perturbations due to these
modes\cite{tamaopen}.  Unfortunately, however, we have no rigorous proof for
it. One possible (and probably most reasonable) stand point is to require the
square integrability of the mode functions, otherwise integration by parts
cannot be performed. Then we find that the discrete modes at $p^2=-4$ should
be excluded from the spectrum. This can be seen as follows.

{}For a while, we neglect a positive definite 
term $\kappa\varphi_0'{}^2/2$ in the operator ${\widehat{\cal O}}$. 
Then Eq.~(\ref{eq:schlike}) for $p^2=-4$ becomes
\begin{equation}
  {\widehat{\cal O}}_0[q]=0,\quad{\widehat{\cal
      O}}_0:=-\frac{d^2}{d\eta_{E}^2} +
  \varphi_0'\left(\frac{1}{\varphi_0'}\right)''\,.
\label{spmode}
\end{equation}
The two independent solutions are easily found as
\begin{equation}
 {q}_1=\frac{N_1}{\varphi_0'}\,,\hskip 1cm
 {q}_2=\frac{N_2}{\varphi_0'}\int_{-\infty}^{\eta_E}\varphi_0'{}^2d\eta_E\,,
 \label{spsol}
\end{equation}
where $N_1$ and $N_2$ are constants. One readily sees that ${q}_1$ is badly
divergent at $\eta_E=\pm\infty$ since $\varphi_0'$ vanishes there (this is
necessary for regularity of the instanton). As for ${q}_2$, it is regular at
$\eta_E=-\infty$ by construction but diverges at $\eta_E=\infty$. Thus there
is no square-integrable solution in the limit of weak gravitational coupling.
Now, since the solution ${q}_2$ has no node, the lowest eigenvalue $p_0^2+4$
of the operator ${\widehat{\cal O}}_0$ will be greater than zero. Then since
the term we have neglected from ${\widehat{\cal O}}$ is positive definite, the
lowest eigenvalue of Eq.~(\ref{eq:schlike}) will be greater than $p_0^2+4$,
which is positive definite. Hence we conclude that no square-integrable
solution of Eq.~(\ref{eq:schlike}) exists at $p^2=-4$.

If the operator ${\widehat{\cal O}}$ in front of $\bbox{q}$ were absent from
the action (\ref{eq:mainoutsc}), the above fact would be sufficient to exclude
the modes with $p^2=-4$. However, the action for these modes seems to be an
indeterminacy of the form zero times infinity. On one hand, ${\widehat{\cal
    O}}$ vanishes, but on the other, $\bbox{q}$ are not integrable on the
spacelike surface $\chi = 0$. As we have mentioned, however, there are
evidences that they do not contribute to physical fluctuations. Hence it seems
reasonable to accept the square integrability as the guiding principle and
exclude these special modes from the spectrum.

For the tensor part, we have
\begin{equation}
\label{eq:mainoutvec}
{\cal S}_{{\bbox{w}}}^{(2)} = 2\kappa \int \frac{\sqrt{-\gamma^{dS}}}{2}\,
({\widehat K}\,{\bbox{w}})\, \left\{\,{^{dS}\Box-1-{\widehat K} \over ^{(2)}
    \triangle(^{(2)}\triangle + 2)} \,\right\} \,{\bbox{w}}\,d\eta_{E}\,d^3x
\end{equation}
where ${\widehat K}$ is the operator defined in Eq.~(\ref{Optensor}).  If we
expand ${\bbox{w}}$ by means of the eigenfunction of ${\widehat K}$, the
operator ${\widehat K}$ can be replaced with the corresponding eigenvalue.
Then we can absorb the factor
\begin{equation}
{2\kappa\, {\widehat K}\over ^{(2)}\triangle(^{(2)}\triangle + 2)}=
\displaystyle{2\kappa\, p^2\over (\ell -1)\ell(\ell +1)(\ell +2)}=:
{1\over ({\cal N}_{+}^{p\ell})^{2}},
\end{equation}
by redefinition of variable and we obtain the action for an ordinary scalar
field.  As before ${\bbox{w}}$ is decomposed as ${\bbox{w}} = a_E \sum {\cal
  N}_{+}^{p\ell} {\bbox{w}}_{p\ell m} U_T^{p\ell m}$.  As in the case of odd
parity, the spectrum is purely continuous, with $p^2>0$.

Now we relate ${\bbox{w}}_{p\ell m}$ with the mode function 
$U^{(+)}_{p\ell m}$\cite{tamaopen} 
defined in the inside of the lightcone (See Appendix B). 
In order to relate the amplitude ${\bbox{w}}_{p\ell m}$ to 
$U^{(+)}_{p\ell m}$, we focus on the traceless part of the 
$(AB)$-component 
of the metric perturbation, $v$, in the synchronous gauge 
($A=S=T=0$). As before, we associate 
a subscript (or superscript) $N$ to indicate the quantity 
evaluated in this gauge. 

{}From Eqs.~(\ref{lcl}), (\ref{defv}) and 
(\ref{defkv}) and with the aid of the equation 
\begin{equation}
  \dot{\bbox{w}}={^{(2)}\triangle(^{(2)}\triangle+2)\over 2\kappa{\widehat K}
    c_{E}\sqrt{w}} \bbox{\Pi}_w,
\end{equation}
which follows from the reduced Lagrangian (\ref{Lagw}), 
$v'_{N}$, is evaluated as 
\begin{eqnarray}
   v'_N = - \frac{\kappa{\cal N}_+^{p\ell}}
        {a_{E}c_{E}^2 {^{(2)}\triangle}({^{
        (2)}\triangle}+2)}\left(2h_{E} \partial_\rho +
    ({^{(2)}\triangle} + 2h_{E}^2 - 2c_{E}^2{\widehat K})
   \right)  {\cal Y}^{p\ell m}{\bbox{w}}_{p\ell m}
      u^{p}  
\end{eqnarray}
On the other hand, 
using the expression for the tensor harmonics 
given in \cite{tomita,tamaopen}, 
after the analytic continuation to region C, we have 
\begin{eqnarray}
 v_N & = & -\sum_{p\ell m} c_{E}^{-2}
       \sqrt{2\Gamma(ip+\ell +1)\Gamma(-ip+\ell +1)\over
       p^2(p^2+1)\Gamma(ip)\Gamma(-ip)}\cr
   &&\quad\times\left(2h_{E}\partial_{\rho}+(-\ell(\ell+1)+2h_{E}^2
     -2p^2 c_{E}^2)\right) {\cal P}_{p\ell}Y^{\ell m}U^{(+)}_{p\ell m}\, .
\end{eqnarray}

Hence, we find that the variable ${\bbox{w}}$ is related to the variable
$U^{(+)}_{p\ell}$ through the
relation
\begin{equation}
  \label{eq:comp}
  {\bbox{w}}_{{p}\ell m}u^p =
  \frac{1}{\sqrt{2\,\kappa}\,i^{\ell+1}}\sqrt{\frac{\Gamma(-ip+\ell+1)}
{(p^2+1)\Gamma(ip+\ell+1)\Gamma(ip)\Gamma(-ip)}}\,a_{E}\,
  {dU^{(+)}_{p\ell}\over d\eta_{E}}\,.
\end{equation}
As in the odd parity case, using the equation for $U^{(+)}$ which is the 
same as for $U^{(-)}$, Eq.~(\ref{Ueq}), the inverse relation is
given by
\begin{equation}
  \label{eq:compp}
 U^{(+)}_{p\ell}=-\sqrt{2\,\kappa}\,i^{\ell+1}\,
\sqrt{\frac{\Gamma(ip+\ell+1)\Gamma(ip)\Gamma(-ip)}
{(p^2+1)\Gamma(-ip+\ell+1)}}\,a_{E}^{-2}{\bbox{w}}_{p\ell m}
  {d(a_{E} u^p)\over d\eta_{E}}\,.
\end{equation}

\section {Spectrum of Gravity Waves}

As an application of our formalism, let us find the spectrum of long
wavelength tensor modes predicted in open inflationary models. This reduces to
solving the scattering problem for the Schr{\"o}dinger equation 
(\ref{upspec}). 
The potential for this problem vanishes at both $\eta_{E}\to -\infty$ and
$\eta_{E}\to\infty$, so the asymptotic behavior at $\pm\infty$ of the two
orthogonal solutions $u_{(\pm)}^{p}$ for the energy $p^2$ can
be taken just as incident plane waves from $\pm\infty$ with momentum $p$ which
interact with the potential and produce reflected waves which return to
$\pm\infty$ with reflection amplitude $\sigma_{ \pm}$, and transmitted waves
moving to $\mp\infty$ with transmission amplitude $\varrho_{ \pm}$.  That is,
in the limit $\eta_{E}\to\pm\infty$, the $u_{(\pm)}^{p}$ are given by
\begin{equation}
 u_{(+)}^{p}= \left\{\begin{array}{ll} 
      \displaystyle \frac{1}{\sqrt{2\pi}} \,\left(
        \varrho_{ +} \,e^{ip\eta_{E}} + e^{-ip\eta_{E}}\right)
      &(\eta_{E}\rightarrow +\infty),\\
      \displaystyle \frac{1}{\sqrt{2\pi}} \, \sigma_{ +} \,e^{-ip\eta_{E}}
      &(\eta_{E}\rightarrow -\infty),
\end{array}\right.
\end{equation}
and
\begin{equation}
u_{(-)}^{p} =\left\{\begin{array}{ll} 
      \displaystyle \frac{1}{\sqrt{2\pi}} \, \sigma_{
        -} \,e^{ip\eta_{E}}
      &(\eta_{E}\rightarrow +\infty),\\
      \displaystyle \frac{1}{\sqrt{2\pi}} \,\left( \varrho_{ -}
        \,e^{-ip\eta_{E}} + e^{ip\eta_{E}}\right)
      &(\eta_{E}\rightarrow -\infty).
    \end{array}\right.
\end{equation} 
Here we note that $p$ is non-negative. 
Using the Wronskian relations we obtain 
\begin{eqnarray}
  && \vert\sigma_{ +}  \vert^2
    =1-\vert\varrho_{ +} \vert^2, 
\quad 
  \vert\sigma_{ -}  \vert^2
    =1-\vert\varrho_{ -} \vert^2, 
\label{Wron1}
\\ 
  && \sigma_{ -} =\sigma_{ +},
\quad \sigma_{ +}  \overline{\rho_{-}} +
\overline{\sigma_{-}}  \rho_{ +} = 0. 
\label{Wron2}
\end{eqnarray}
Using Eq.~(\ref{Wron2}), we can show that 
\begin{equation}
 \int d\eta_{E}\, u_{(+)}^{p}
   \overline{u_{(-)}^{p}} =0, 
\end{equation}
and hence $u_{(+)}^{p}$ and 
$u_{(-)}^{p}$ are orthogonal. 
Note that normalization condition (\ref{upnorm}) is satisfied because
\begin{equation}
 {1\over 2}
 \left(\vert\sigma_{ \pm}  \vert^2
    +\vert\varrho_{ \pm} \vert^2+1\right)=1. 
\end{equation} 

Analytically continuing the solution inside of the lightcone by means of
$\eta_{E} = -\eta_R - i\pi/2$, where
$\eta_R$ is the conformal time in region R, the amplitude
of perturbations well after the modes have crossed the horizon
($\eta_{ R}\to 0 $) is given by
\begin{eqnarray}
  |u_{(+)}^{p}|^2 + |u_{(-)}^{p}|^2 
   &=& \vert e^{\pi p/2}\sigma_{ -}\vert^2
       +\vert e^{\pi p/2}\varrho_{ +}
       +e^{-\pi p/2}\vert^2
 \nonumber\\
  &=&   \frac{1}{\pi} \left(\cosh \pi p +
\Re\,\varrho_{ +}\right).
\label{general}
\end{eqnarray}
As we can see from the equation above, the bubble manifests itself in the
spectrum through the real part of the reflection amplitude of the
Schr{\"o}dinger problem.

In the thin-wall approximation, we can take the interior of the bubble as a
pure de Sitter space, with scale factor given by $a_{E} = 1/(H \cosh \eta_{E})$
outside the lightcone from N.  In this limit, we can integrate out the
potential in (\ref{upspec}) and express it as a delta function with strength
$\Delta s = \kappa \mu R_W/2$, where $\mu=\int a_{E}^{-1}
\varphi'_{0}{}^2\,d\eta_{E}$ and $R_W = a_{E}(\eta_w)$ are the surface tension
and the radius of the wall, respectively. The reflection amplitude for this
delta function potential is $\varrho_{ +} = - i e^{-2ip\eta_w}\Delta
s/(2p+i\,\Delta s)$. Using Eq.~(\ref{eq:compp}) with the fact that the scale
factor inside the lightcone is given by $a=-1/(H\sinh \eta_R)$, we recover
the spectrum for $U^{(+)}_{p\ell}$ given in reference \cite{tamaanis}:
\begin{eqnarray}
   \langle U^{{(+)}2}_{p\ell}\rangle
&=&{2\pi \kappa H^2\over p(p^2+1)\sinh\pi p}\left( 
  |u_{(+)}^{p}|^2 + |u_{(-)}^{p}|^2 \right) \nonumber\\
&=& \frac{2 \kappa H^2 \coth \pi p}{p(p^2 +1)}\left(1
- \frac{R}{\cosh \pi p }\left(\frac{\Delta s \cos 2 p\eta_w+ 2 p\sin
2p\eta_w}{\Delta s}\right)\right),
\end{eqnarray}
where $R$ is the reflection coefficient, given
by
\begin{equation}
  R = \frac{(\Delta s)^2}{4p^2 +(\Delta s)^2}.
\end{equation}

Of course, the validity of Eq.(\ref{general}) is not restricted to the thin
wall regime, and a more complete analysis in the general case will be
presented elsewhere \cite{stspec}. Notice that this equation has been derived
neglecting the term $\kappa\varphi_0'{}^2/2$ in the equation for the
evolution of the modes inside the lightcone. This is justified for long
wavelength modes, which are frozen in soon after bubble nucleation.  For modes
that enter the horizon at times $t>>H^{-1}$, which corresponds to $p>>1$, the
generation of perturbations occurs during the second stage of inflation and
has little to do with the bubble profile. Instead, the details of the slow
roll potential will be important.

\section{Conclusions and discussion}

In this paper we have applied Fadeev and Jackiw's formalism for constrained
systems to the problem of cosmological perturbations in the one bubble open
inflationary universe (The cases of flat and closed universes have been also
considered in Appendix \ref{sec:redins}).

We have found the reduced action for a gauge invariant variable
describing scalar degrees of freedom and for two gauge invariant
variables describing tensor degrees of freedom. 
This tensor part coincides with the one found previously in
\cite{tamaopen}.

The nucleation of a bubble breaks the $O(4,1)$ symmetry of de Sitter
space down to $O(3,1)$. It is known that, neglecting the self-gravity of
the bubble, there is a special scalar mode with eigenvalue $p^2=-4$
which corresponds to fluctuations of the bubble wall. This mode can be
seen as the Goldstone mode associated with the breaking of symmetry
$O(4,1)$ down to $O(3,1)$.  We have seen that the wall fluctuation mode 
disappears from the spectrum of scalar perturbations once gravity is 
included.  This is somewhat reminiscent of what
happens in gauge theories: the Goldstone mode disappears when the gauge fields
(gravitational degrees of freedom) are included (the tensor modes acquire
a mass term on the wall, where the Goldstone used to live, which cuts off the
infrared divergence encountered in \cite{allen}). As pointed out in
\cite{tamaanis} (see also \cite{ishish}), 
the disappearance of the wall fluctuation mode is perhaps not
too surprising.  Even in the absence of self-gravity, this mode can be written
as a tensor mode, which contributes to microwave background anisotropies just
like any gravitational wave would do \cite{k3,bell2,jaume}.  
The study of tensor modes in \cite{tamaopen,tamaanis} showed that in the 
weak gravity limit, the `infrared' contribution of gravity waves to 
microwave anisotropies reproduces the effect of bubble wall fluctuations.

As an application of our formalism, we have derived the spectrum of long
wavelength tensor modes in open inflation. This is given in terms of the real
part of the reflection coefficient in a one dimensional scattering problem,
where the potential barrier is a function of the bubble profile. In the thin
wall regime, we recover the results of \cite{tamaopen}. A more
complete study of this spectrum and that of scalar perturbations will be
presented elsewhere\cite{stspec}.

\section*{acknowledgements}

J.G. and X.M. acknowledge financial support from CICYT under contract
AEN95-0882 and from European Project CI1-CT94-0004. This work was
supported in part by Monbusho Grant-in-Aid for Scientific Research
No. 09640355.

\appendix
\section{Second Order Lagrangian}
\label{sec:seconorderin}
In this appendix we derive the action for the fluctuations of a scalar field
coupled to gravity on a FRW background, up to second order in the perturbation
variables.

The metric is written in the ADM \cite{ADM} form
\begin{equation}
  ds^2 = -({\cal N}^2 - {\cal N}_i {\cal N}^i) d\eta^2\, +\, 2 {\cal N}_i dx^i
  d\eta\, + \,^{(3)}g_{ij} dx^i dx^j,
\end{equation}
where $\cal N$ is the lapse, ${\cal N}_i$ is the shift function and
$^{(3)}g_{ij}$ is the metric on the constant $\eta$
space-like hypersurfaces in which we have foliated spacetime. Up to total
derivative terms, the purely gravitational part of the action can be written
as
\begin{eqnarray}
  {\cal S}_{gr}&=& \frac{1}{2 \kappa} \int d^4x \sqrt{-^{(4)}g}\,^{(4)}R
  = \frac{1}{2 \kappa} \int d^4x\{{\cal N}\sqrt{^{(3)}g}(K^{ij}K_{ij} -
  K^i{}_i K^j{}_j) \\
  & & + \frac{1}{2}\partial_i(\sqrt{^{(3)}g}{\cal N} \,^{
    (3)}g^{ij}) \partial_j \ln\, ^{(3)}g + \partial_i{\cal
    N}\partial_j(\sqrt{^{(3)}g}\,{}^{(3)}g^{ij}) - \frac{1}{2} {\cal
    N}\sqrt{^{(3)}g}\,^3\Gamma^l{}_{ij}\partial_l{}^{(3)}g^{ij}\},\nonumber
\end{eqnarray}
where
\begin{equation}
  K_{ij} = \frac{1}{2 {\cal N}}({\cal N}_{i|j} + {\cal N}_{j|i}
  -\,^{(3)}g'_{ij}).
\end{equation}
With a prime we denote a derivative with respect to time $\eta$, and the
vertical bar stands for the covariant derivative with respect to the spatial
metric $^{(3)}g_{ij}$. The action for the scalar matter
field is
\begin{equation}
  {\cal S}_m =\int d^4x \sqrt{-{}^{(4)}g}\,
\left\{-\frac{1}{2}\nabla_\mu \varphi\nabla^\mu\varphi - V(\varphi)\right\}.
\end{equation}

Now we expand the metric and the scalar field over an FRW-like background
solution. The perturbed metric and the perturbed scalar field read
\begin{equation}
\begin{array}{lcl}
  ds^2 & = & a(\eta)^2\{- (1 + 2A)d\eta^2 + 2 S_{i} dx^i d\eta +
  (\gamma_{ij} + h_{ij})dx^i dx^j\} \nonumber,\\ 
  \varphi & = & \varphi_0(\eta) + \delta\varphi,\label{eq:perturb}
\end{array}
\end{equation}
where $\gamma_{ij}$ is the metric on the constant curvature space
sections. The background fields $a$ and $\varphi_0$ satisfy
the equations
\begin{equation}
  \begin{array}{c} {\cal H}^2 - {\cal H}' + {\cal K} = 
    \displaystyle\frac{\kappa}{2}
    \,\varphi_0{}'{}^2 ,\\
    2\, {\cal H}' + {\cal H}^2 + {\cal K} = \displaystyle\frac{\kappa}{2} (-
    \varphi_0'^2 + 2\, a^2 V(\varphi_0)), \\
    \varphi_0{}'' + 2\, {\cal H} \varphi_0{}' + a^2 V_{,\varphi}(\varphi_0) =
    0, \end{array}
\end{equation}
where ${\cal H} := a'/a$, and $\cal K$ is the curvature parameter, which
has the values 1, 0, -1 for closed, flat and open universes respectively.

Expanding the total action, keeping terms of second order in perturbations,
and using the background equations, we find
\begin{equation}
  {\cal S} = {\cal S}_{gr} + {\cal S}_m = {\cal S}_0 + \delta_2{\cal S},
\end{equation}
where ${\cal S}_0$ is the action for the background solution and
$\delta_2{\cal S}$ is quadratic in perturbations:
\begin{eqnarray}
  \label{eq:totalaction} \delta_2{\cal S} &=& \frac{1}{2 \kappa} \int d^4x
  \,a^2 \sqrt{\gamma}\{-2(2 {\cal H}^2 + {\cal H}') A^2 + (S_{(i|j)} -
  \frac{h'_{ij}}{2})^2 - (S^i{}_{|i} - \frac{h^i{}_{i}{}'}{2})^2\nonumber \\ &
  & \hskip 1 cm +\, \frac{1}{4}(2 h^{ij|k}h_{jk|i} - h^{ij|k}h_{ij|k} - 2
  h^{ij}{}_{|j} h^k{}_{k|i} + h^i{}_i{}^{|j}h^k{}_{k|j}) + A^{|i}(h^j{}_{j|i}
  - h^j{}_{i|j}) \nonumber\\ & & \hskip 1 cm +\, 2
  \frac{\kappa}{2}(\delta\varphi'^2 - \delta\varphi^{|i}\delta\varphi_{|i} -
  a^2 V_{,\varphi\varphi} \delta\varphi^2) - 2 \kappa(\varphi_0'
  \delta\varphi' A + a^2 V_{,\varphi} \delta\varphi A)\nonumber\\ & & \hskip 1
  cm +\, 4 (S^i{}_{|i} - \frac{h^i{}_{i}{}'}{2})(\frac{\kappa}{2} \varphi_0'
  \delta\varphi - {\cal H} A) + {\cal K}(2 A^2 - 2 A h^i{}_i - \frac{1}{2}
  h^i{}_i h^j{}_j + h^{ij} h_{ij})\}.
\end{eqnarray}
We have raised and lowered spatial indices with $\gamma_{ij}$.

In the open inflation case, comparing (\ref{eq:perturb}) with
(\ref{eq:linepertout}), we see that the action for small perturbations outside
the lightcone can be found just replacing $a^2\rightarrow -a_{E}^2$ and
$\gamma_{ij} \rightarrow - \gamma_{ij}^{dS}$ in expression
(\ref{eq:totalaction}). The result is equation (\ref{eq:totalactioncont}) in
the text.

\section{Reduction inside the lightcone}
\label{sec:redins}

Although in the case of open inflation the $t=const.$ surfaces are not Cauchy
surfaces for the whole spacetime, it is known \cite{tamaopen} that they can be
used to normalize the subcurvature modes, i.e. those modes for which the
eigenvalue of the Laplacian on the hyperboloids of homogeneity and isotropy is
smaller than -1.  Furthermore, the resulting reduced action can be, in some
heuristic sense to be discussed later, analytically continued to the outside
of the lightcone.  Then it is found that we obtain the correct result even for
supercurvature modes, i.e. modes other than the subcurvature modes.  Compared
with the reduction in the outside of the lightcone, the analysis in the inside
of the lightcone is very simple.  Therefore, for an alternative less rigorous
but rapid derivation, in this appendix we consider the reduction of the
Lagrangian directly inside the lightcone.  We shall simultaneously consider
also the case of flat and closed spatial sections.

Inside the lightcone the metric perturbations can be decomposed into scalar,
vector and tensor modes\cite{bardeen,kodsas,branden}, regarding the way
in which the
modes transform under spatial coordinate transformations. On a homogeneous
background, the modes are decoupled in the action, and evolve
independently. Thus we expand the metric perturbations as
\begin{equation}
\begin{array}{lcl}
  \label{eq:decomposition} 
  h_{ij} &=& -2 \psi \gamma_{ij} + 2 E_{|ij} + 2
  F_{(i|j)} + t_{ij}, \\
  S_i &=& B_{|i} + V_i,
\end{array}
\end{equation}
where $\psi$, $B$ and $E$ are scalar modes, $F_i$ and $V_i$ are vector
modes,\footnote{It should be noted that $F_i$ and $V_i$ defined here are 
different from $F_A$ and $V_A$ defined in Eq.~(\ref{eq:expout}).}
and $t_{ij}$ is a tensor mode. $F_i$ and $V_i$ are divergenceless, and
$t_{ij}$ is transverse traceless (${TT}$), i.e.
\begin{equation}
  F^i{}_{|i} = V^i{}_{|i} = 0 = t^i{}_i = t^{ij}{}_{|j}.
\end{equation}

Substituting the decomposition (\ref{eq:decomposition}) in
(\ref{eq:totalaction}), the action is decoupled into three pieces
\begin{equation}
  \delta_2{\cal S} = ^{(s)}\delta_2{\cal S} + ^{(v)}\delta_2{\cal S} +
  ^{(t)}\delta_2{\cal S}.
\end{equation}

\subsection{Scalar Perturbations}
The action for scalar perturbations reads
\begin{eqnarray}
  \label{eq:scalaraction} ^{(s)}\delta_2{\cal S} &=& \frac{1}{2 \kappa} \int
  d^4x \,a^2\sqrt{\gamma}\{-6 \psi'^2 - 12 {\cal H} A \psi' + 2 \triangle
  \psi(2 A - \psi) - 2 ({\cal H}' + 2 {\cal H}^2) A^2 \nonumber \\ & & +
  \kappa (\delta\varphi'^2 + \delta\varphi \triangle\delta\varphi - a^2
  V_{,\varphi\varphi} \delta\varphi^2) + 2 \kappa (3 \varphi_0' \psi'
  \delta\varphi - \varphi_0' \delta\varphi' A - a^2 V_{,\varphi}
  A\delta\varphi) \nonumber \\ & & + {\cal K} (-6 \psi^2 + 2 A^2
  + 12 \psi A + 2 (B-E') \triangle (B-E'))\} \nonumber\\
  & &+ 4 \triangle (B-E') (\frac{\kappa}{2}
  \varphi_0'\delta\varphi - \psi' - {\cal H} A) ,
\end{eqnarray}
where
$\triangle$ is the laplacian associated with $\gamma_{ij}$.

To apply FJ formalism\cite{tears}, we first have to cast the Lagrangian in
first order form, defining momenta for the variables whose time derivative
appears quadratically in the Lagrangian.  
As usual, the conjugate momenta are defined as
\begin{equation}
  \label{defppsi}
\begin{array}{lcl}
  \Pi_\psi &:=&\displaystyle \frac{\delta }{\delta \dot \psi}
  {}^{(s)}\delta_2{\cal S}=
  \frac{2\,a^2\,\sqrt{\gamma}}{\kappa}\left(-3\,\psi' + \triangle E' +
    3\,\frac{\kappa}{2}\varphi_0'\,\delta\varphi - \triangle B - 3\,{\cal
      H}\, A
  \right),\\
  \Pi_{\delta\varphi} &:=& \displaystyle \frac{\delta }{\delta \,\delta
    \dot\varphi }{}^{(s)}\delta_2{\cal S}=
  a^2\,\sqrt{\gamma}\left(\delta\varphi' -
    \varphi_0'\, A\right),\\
  \Pi_E &:=& \displaystyle \frac{\delta }{\delta \dot
    E}{}^{(s)}\delta_2{\cal S}=
  \frac{2\,a^2\,\sqrt{\gamma}\,\triangle}{\kappa}\left({\cal K}\,E'
    +\psi'-\frac{\kappa}{2}\varphi_0'\,\delta\varphi- {\cal K}\,B +{\cal
      H}\, A\right).
\end{array}
\end{equation}
The first order Lagrangian for
scalar perturbations turns out to be
\begin{eqnarray}
  \label{eq:lagrangscalar} ^{(s)}{\cal L} &=& ^{(s)}{\cal L}_1 + ^{(s)}{\cal
    L}_0 = ^{(s)}{\cal L}_1 - ^{(s)}{\cal H} - B\,{\cal C}_B - A\,{\cal
    C}_{A}, \\
  ^{(s)}{\cal L}_1 &=& \Pi_\psi \psi' +
  \Pi_{\delta\varphi}\delta\varphi' + \Pi_E E'\nonumber ,\\
  ^{(s)}{\cal H} &=& \frac{\kappa}{4\,a^2\sqrt{\gamma}(\triangle + 3\,{\cal
      K})}\left( - {\cal K}\,\Pi_\psi^2 + 2\,\Pi_\psi \Pi_E + \frac{3
      \Pi_E^2}{\triangle} + 2(\triangle + 3\,{\cal
      K})\frac{\Pi_{\delta\varphi}^2}{\kappa}\right) +
  \frac{\kappa}{2} \varphi_0' \,\Pi_\psi\delta\varphi \nonumber\\
  & &+\, \frac{a^2\sqrt{\gamma}}{\kappa}\left((\triangle + 3\,{\cal
      K})\psi^2 - \frac{\kappa}{2}\,\left( \triangle + 3\,{\cal K} - {\cal
        H}^2 - {\cal H}'
      +\frac{\varphi_0'''}{\varphi_0'}\right)\delta\varphi^2\right),\nonumber\\
  {\cal C}_B &=& \Pi_E \nonumber ,\\
  {\cal C}_A &=& -{\cal H} \Pi_\psi + \varphi_0'\Pi_{\delta\varphi} +
  \frac{2\,a^2\sqrt{\gamma}}{\kappa}\left(-( \triangle + 3\,{\cal K}) \psi
    + \frac{\kappa}{2} ({\cal H}\varphi_0' - \varphi_0'')
    \delta\varphi\right).\nonumber
\end{eqnarray}

We observe that neither $A$ nor $B$ enters into ${\cal L}_1$, so there is no
dynamical evolution for these fields. They correspond to $\delta{\cal
  N}^{\mu}$ in the notation of the introduction.  These fields appear linearly
in the lagrangian, and their equations of motion, ${\cal C}_{A/B}=0$ contain
no time derivatives. They allow us to evaluate two of the momenta in terms of
the other fields. Moreover, the constraints ${\cal C}_A$ and ${\cal C}_B$ are
the generators of the infinitesimal gauge transformations associated with
diffeomorphisms.  Under a scalar diffeomorphism generated by the vector
$\lambda^\mu = (\lambda^0,\lambda^{|i})$, the metric transforms into
$g_{\mu\nu} + \delta_g g_{\mu\nu}^0$, from which we can read the
variation of all the scalar components of the metric perturbation. By
commutation with $\lambda^0\,{\cal C}_A + \lambda\,{\cal C}_B$ we recover the
transformation law for the canonical fields:
\begin{equation}
  \begin{array}{rclcrcl} \delta_g\psi &=& - {\cal H}\,\lambda^0, &\hskip
    1cm & \delta_g \Pi_\psi
    &=&\,\displaystyle{\frac{2\,a^2\sqrt{\gamma}}{\kappa}}
    (\triangle +3\,{\cal
      K})\,\lambda^0,\\
    \delta_g\delta\varphi &=& \varphi_0'\, \lambda^0,& & \delta_g
    \Pi_{\delta\varphi} &=& a^2\sqrt{\gamma}( \varphi_0'' -
    \varphi_0' {\cal H})\,\lambda^0,\\
    \delta_g E &=& \lambda ,& \hskip 1cm &\delta_g \Pi_E &=& 0.
  \end{array}
\label{gtf}
\end{equation} 
The constraints and the scalar Hamiltonian $^{(s)}{\cal H}$ satisfy the
following algebra:
\begin{eqnarray}
  \{^{(s)}{\cal H},{\cal C}_A\} = \partial_\eta {\cal C}_A - {\cal H}\,{\cal
    C}_A + {\cal C}_B , \hskip 1cm \{^{(s)}{\cal H},{\cal C}_B\} = 0, \hskip
  1cm \{{\cal C}_A,{\cal C}_B\} = 0.
\end{eqnarray}
The time derivative of the constraints appears due to its explicit time
dependence. This derivative acts only on background quantities but not on
canonical coordinates.  Under a gauge transformation, $^{(s)}{\cal L}_1$
transforms as
\begin{equation}
  \delta_g{}^{(s)}{\cal
  L}_1=-\lambda^0
\left(\frac{d}{d\eta}-\frac{\partial}{\partial\eta}\right){\cal C}_A
-\lambda
\left(\frac{d}{d\eta}-\frac{\partial}{\partial\eta}\right){\cal C}_B \,.
\end{equation}
Using these results and the fact that the action is
invariant under a gauge transformation, we can recover the transformation law
for the lagrange multipliers \cite{wipf}:
\begin{equation}
  \delta_g A = \lambda^0{}' + {\cal H}\lambda^0,\hskip 2 cm
  \delta_g B = \lambda' - \lambda^0.
\end{equation}

Now we proceed with the phase space reduction. We start by solving the
constraints ${\cal C}_B=0$ and ${\cal C}_{A} = 0$ for $\Pi_E$ and
$\Pi_{\delta\varphi}$, respectively.  After substitution of the constraint,
there is no $E$ dependence in the Lagrangian, so the Lagrangian becomes a
functional of only $\Pi_\psi$, $\psi$ and $\delta\varphi$, $^{(s)}{\cal L}^* =
\,{}^{(s)}{\cal L}^*[\Pi_\psi,\psi,\delta\varphi]$.  The dissaperance of $E$
is related to the fact that there is no variable other than $E$ itself whose
gauge transformation depends on $\lambda$ besides the Lagrange multipliers.
Hence it is not possible to construct a gauge invariant combination which
contains $E$.  Therefore $E$ necessarily vanishes from $^{(s)}{\cal L}^{*}$.

Applying the formula given in (\ref{giv}) or equivalently 
looking at the gauge transformation law given in Eqs~(\ref{gtf}), 
the gauge invariant combinations 
are found to be constructed from the remaining variables as 
\begin{equation}
\begin{array}{lcl}
  {\bf \Psi} & = & \displaystyle \psi + 
  \frac{{\cal H}}{\varphi_0'}\delta\varphi,\\
  {\bf \Pi}_{\bf \Psi} & = & \displaystyle \Pi_\psi - \frac{2\,a^2
    \sqrt{\gamma}}{\kappa \varphi_0'}(\triangle+3\,{\cal K})\delta\varphi.
\end{array}
\end{equation}
The action expressed as a functional of ${\bf \Pi}_{\bf \Psi}$ and
 ${\bf \Psi}$ can be obtained, up to total derivative terms, by simply
 putting $\delta\varphi = 0$ in
  $^{(s)}{\cal L}^*[\Pi_\psi,\psi,\delta\varphi]$ and replacing
 $\Pi_\psi$ with ${\bf \Pi}_{\bf \Psi}$ and $\psi$ with ${\bf \Psi}$. 
We finally obtain
\begin{eqnarray}
  ^{(s)}{\cal L}^* & = & {\bf \Pi}_{\bf \Psi} {\bf \Psi}' - \frac{2\,a^2
    \sqrt{\gamma}}{\kappa^2 \varphi'_0{}^2} ((\triangle + 3\,{\cal K})
  {\bf \Psi} + \frac{\kappa {\cal H}}{2\,a^2 \sqrt{\gamma}} {\bf
    \Pi}_{\bf \Psi})^2
  \nonumber \\
  & & - \frac{a^2 \sqrt{\gamma}}{\kappa}{\bf \Psi}(\triangle + 3\,{\cal
    K}){\bf \Psi} + \frac{1}{4 a^2 \sqrt{\gamma}}{\bf \Pi}_{\bf \Psi}
  \frac{\kappa {\cal K}}{(\triangle + 3 {\cal K})}{\bf \Pi}_{\bf \Psi}.
\label{B12}
\end{eqnarray}
Notice that the procedure is equivalent to fixing a gauge
where $E$ and $\delta\varphi$ (the variables conjugate to the momenta we have
solved the constraints for) are set to zero.

For flat universes we can easily recover the results of \cite{qflat} or
\cite{branden,mukhanov}.  If ${\cal K}=0$, using the equation of motion for
${\bf \Pi}_{\bf \Psi}$ we can eliminate the momenta ${\bf \Pi}_{\bf
  \Psi}$ in favor of the velocity ${\bf \Psi}'$, and we will find the
following second order Lagrangian:
\begin{equation}
  {\cal L}_{{\bf \Psi}} = \frac{1}{2}\frac{a^2\varphi_0'^2}{{\cal
      H}^2}({\bf \Psi}'^2 + {\bf \Psi} \triangle {\bf \Psi}),
\end{equation}
which after the rescaling
\begin{equation}
  \bbox{\vartheta} = z \,{\bf \Psi}, \hskip 2cm z = \frac{a \varphi_0'}{{\cal
      H}},
\end{equation}
 becomes the Lagrangian of a scalar field in flat spacetime with time
 dependent mass,
\begin{equation}
  {\cal L}_{\bbox{\vartheta}}^{(2)} = \frac{1}{2} \left(\bbox{\vartheta}'^2 -
    \bbox{\vartheta}_{,i} \bbox{\vartheta}^{,i} + \frac{z''}{z} {\bf
      \vartheta}^2\right).
\end{equation}
The reduced gauge invariant variable $\bbox{\vartheta}$ coincides with the one
found in ref\cite{mukhanov}:
\begin{equation}
  \bbox{\vartheta} = a (\delta\varphi + \frac{\varphi_0'}{{\cal H}} \psi).
\end{equation}

In the general case we can perform the canonical transformation
\begin{equation}
\begin{array}{lcl}
  {\bf \Psi} &=& \displaystyle \frac{\kappa\varphi_0'}{4}
 {\bbox{\tilde q}} - \frac{2\,\kappa{\cal
      H}}{a^2\sqrt{\gamma}\varphi_0'}\frac{1}{\triangle + 3\,{\cal K}}\,
  {\bbox{\tilde p}}
  ,\\
  {\bf \Pi}_{\bf \Psi} &=& \displaystyle
  \frac{a^2\sqrt{\gamma}\varphi_0'}{2\,{\cal H}}(\triangle + 3\,{\cal
    K})\,{\bbox{\tilde q}} + \frac{2}{\kappa\varphi_0'}{\bbox{\tilde p}},
\end{array}
\end{equation}
and solve for the momenta ${\bbox{\tilde p}}$ to obtain the second order
Lagrangian
\begin{equation}
\label{eq:mainresult} 
\begin{array}{c}
  \displaystyle {\cal L}^{(2)}_{{\bbox{\tilde q}}} = -\frac{\sqrt{-^{
        (4)}g}}{2}(\triangle + 3\,{\cal K})\,{\bbox{\tilde q}}\left(\Box -
    m^2[a,\varphi_0]\right){\bbox{\tilde q}},
\\
  \displaystyle m^2[a,\varphi_0] = -\frac{1}{a^2}\left(4 \,{\cal K} - 2 {\cal
      H}' + \varphi_0'\left(\frac{1}{\varphi_0'}\right)'' \right).
\end{array}
\end{equation}
Here $\Box$ stands for the four dimensional scalar d'Alembertian. The action
for this lagrangian can also be written as
\begin{equation}
\begin{array}{lcl}
  \label{eq:mainresult2}
  &\displaystyle {\cal S}^{(2)}_{{\bbox{q}_{\bf in}}} =
  \frac{1}{2}\int\,\sqrt{\gamma}(\triangle + 3\,{\cal K}) {{\bbox{q}_{\bf
        in}}} \left({\widehat{\cal O}} + \triangle + 3\,{\cal K}\right)
  {{\bbox{q}_{\bf in}}}\,d\eta d^3x&,\\
  &\displaystyle {\widehat{\cal O}} = - \frac{d^2}{d\eta^2} + \frac{\kappa}{2}
  \varphi_0'{}^2 + \varphi_0'\left(\frac{1}{\varphi_0'}\right)'',&\nonumber
\end{array}
\end{equation}
where ${\bbox{q}_{\bf in}} = a\,{\bbox{\tilde q}}$, which is formally
analogous to (\ref{eq:mainoutsc}). Note, however, that $({\widehat{\cal
    O}}{\bbox{q}})$ is changed by $ (\triangle - 3){\bbox{q}}_{\bf in}$. These
operators are not the analytic continuation of one another, but both have the
same eigenvalues on solutions of the equations of motion.

We note that the reduced gauge invariant variable ${\bbox{\tilde q}}$ is
proportional to ${\bf \Phi}_{\bf H}$ of Bardeen\cite{bardeen},
\begin{equation}
  \label{eq:tildeqin}
  {\bbox{\tilde q}} = \frac{2}{\kappa\varphi_0'}\,\left({\bf \Psi} -
  \frac{{\cal H}\kappa}{2\,a^2\sqrt{\gamma}(\triangle + 3\,{\cal
  K})}\,{\bf \Pi}_{\bf \Psi}
  \right) = \frac{2}{\kappa\varphi_0'}\,(\psi - {\cal
    H}(B-E')) =: -\frac{2}{\kappa\varphi_0'}\,{\bf \Phi}_{\bf H}.
\end{equation}
For flat universes (${\cal K}=0$), ${\bbox{\tilde q}}$ is proportional
to $Q$ of ref \cite{qflat},
\begin{equation}
 {\bbox{\tilde q}} = \frac{Q}{\sqrt{-\triangle}},
\end{equation}
and the action (\ref{eq:mainresult}) reduces to the one found in
ref.~\cite{qflat} if we replace ${\bbox{\tilde q}}$ with $Q$. 
If we take an ansatz of the form ${\bbox{\tilde q}}
=a^{-1}\,{\bbox{q}}^p(\eta)\,Y^{p\ell m}(x^i)$, where $Y^{p\ell m}$ 
are the scalar harmonics on the homogenious spatial section, 
the equation of motion separates into
\begin{equation}
\begin{array}{lcl}
  \triangle\,Y^{p\ell m} &=& (-p^2+{\cal K})\, Y^{p\ell m} ,\\
  {\widehat{\cal O}}[{\bbox{q}}^p]  &=& (p^2 - 4 {\cal K})\,{\bbox{q}}^{p}.
\end{array}
\end{equation}

\subsection{Vector and Tensor Perturbations}

As we have said, neither tensor nor vector modes couple to the scalar
perturbations. The vector part of the action carries, as we will see, no
dynamics. We find for it
\begin{equation}
^{(v)}\delta_2{\cal S} = - \frac{1}{4\kappa}\int \,d^4x
\,a^2\sqrt{\gamma}\left(\tilde V_m - \tilde F_m'\right)^2,
\end{equation}
where
\begin{equation}
  \tilde V_m = \sqrt {-\triangle -2 {\cal K}} \,V_m ,\hskip 2cm \tilde F_m =
  \sqrt {-\triangle -2 {\cal K}} \,F_m.
\end{equation}
We can compute the corresponding first order Lagrangian,
\begin{equation}
  ^{(v)}{\cal L} = \tilde \pi^m \tilde F_m' - \frac{\kappa}{a^2\sqrt{\gamma}}
  \tilde \pi^m \tilde \pi_m + \tilde \pi^m \tilde V_m.  \end{equation} The
field $\tilde V_m$ has no conjugate momenta, and enters as a Lagrange
multiplier which enforces $\tilde \pi^m$ to vanish. After substituting $\tilde
\pi^m = 0$, we will end with a vanishing Lagrangian. In this model vector
modes are pure gauge.

For the tensor modes we find
\begin{equation}
  ^{(t)}\delta_2{\cal S} = \frac{1}{8\kappa}\int \,d^4x
  \,a^2\sqrt{\gamma}\left\{t^{ij}{}'t'_{ij} - t^{ij|k} t_{ij|k} - 2 {\cal K}
  t^{ij}t_{ij}\right\}.
\end{equation}
The action can be cast easily in a first order form,
\begin{equation}
  ^{(t)}\delta_2{\cal S} = \int d^4x \left\{\pi^{ij}t'_{ij} 
   -\frac{2\kappa}{a^2\sqrt{\gamma}}\,\pi^{ij}\pi_{ij} +
    \frac{a^2\sqrt{\gamma}}{8\kappa}
    (t^{ij} (\triangle - 2{\cal K})t_{ij})\right\}.
\end{equation}
The Lagrangian is already in a canonical form and has no constraints. 
We can decompose $t_{ij}$ by using the normalized 
transverse-traceless tensor harmonics\cite{tomita,tamaopen}, 
$Y_{ij}^{(+)p\ell m}$ and $Y_{ij}^{(-)p\ell m}$, as 
\begin{equation}
 t_{ij}=\sum_{p\ell m}U_{p\ell m}^{(+)}(\eta) Y_{ij}^{(+)p\ell m} 
        +\sum_{p\ell m}U_{p\ell m}^{(-)}(\eta) Y_{ij}^{(-)p\ell m}. 
\end{equation}
Then, it becomes manifest that there are 
two decomposed degrees of freedom for each $p,\ell,m$, 
which correspond to gravitational waves.
  
\section{Even parity phase space reduction outside the lightcone}
\label{sec:FJscalar}
Inserting the mode decomposition (\ref{eq:expout}) in the action
(\ref{eq:totalactioncont}), the even parity modes 
decouple from the odd parity ones. The even parity part reads
\begin{eqnarray}
   ^{(s)}{\cal L} &=& \frac{a_{E}^2 c_{E}\sqrt{\omega}}{2 \kappa}\left\{
   S\,\left( {^{(2)}\triangle}\,\xi'{} + 2\,{h_{E}}\,\left(
   w'{} + 2\,\zeta' \right) - 2\,\kappa\varphi_0'\,\delta\dot\varphi +
   4\,{\cal H}\,\dot A - 4\,{\cal H}\,\dot w -
   {\frac{{^{(2)}\triangle}\,\dot T}{{{c_{E}}^2}}} \right)
   \right. \nonumber\\ 
   & &+\, \left( -2 - {\frac{{^{
   (2)}\triangle}}{2\,{{c_{E}}^2}}} \right) \,{{S}^2} - 2\,S'{}\,\dot w +
   {\frac{\left( -{^{(2)}\triangle} -
   2\,{^{(2)}\triangle}\,{{h_{E}}^2} \right) \,
   {{T}^2}}{{{c_{E}}^2}}} - {^{(2)}\triangle}\,\xi'{}\,\dot T
   - {\frac{{^{(2)}\triangle}\,{{\dot T}^2}}{2\,{{c_{E}}^2}}}\nonumber\\
   & & +  \,T\,\left( -4\,{^{(2)}\triangle}\,{\cal
   H}\, A + 2\,\kappa\,{^{(2)}\triangle}\,\delta\varphi\,
   \varphi_0' - {}{^{ {(2)}}\triangle} \left( 2 + {^{(2)}\triangle} \right)
   \,v' - {^{(2)}\triangle}\,w'{} +
   2\,{^{(2)}\triangle}\,{h_{E}}\,\xi'{} \right. \nonumber\\ 
   & & + \,2\,{^{(2)}\triangle}\,\zeta'{} +\,\left.
   {\frac{2\,{^{(2)}\triangle}\,{h_{E}}\,\dot T}{{{c_{E}}^2}}}
   \right) - 2\,h_{E}^2 \,{{\zeta}^2} +
   2\,{{c_{E}}^2}\,\left( -2\,{\cal H}\, A +
   \kappa\varphi_0'\,\delta\varphi - w'{} \right) \, \zeta'{} 
   -\,{\frac{{{\dot w}^2}}{2}}
   \nonumber\\
   & & +\,\zeta\, \left( \left( 2 +
   {^{(2)}\triangle} \right) (w + {}{^{ {(2)}}\triangle}v) 
   - 2\, A\,\left(
   {^{(2)}\triangle} + 2\,{{c_{E}}^2} \right) -
   2\,{^{(2)}\triangle}\,{h_{E}}\,\xi + 4\,{h_{E}}\,\dot A \right. \nonumber\\
   & &\left.
   - 2\,{h_{E}}\,\dot w \right)+\,{^{
   (2)}\triangle}\,{{\xi}^2} - {\frac{{^{
   (2)}\triangle}\,{{c_{E}}^2}\,{{\xi'{}}^2}}{2}} + \xi\,\left(
   -2\,{^{(2)}\triangle}\,{h_{E}}\, A -  {}{^{ {(2)}}\triangle}\left( 2 +
   {^{(2)}\triangle} \right) \dot v  
   \right.\nonumber\\ 
   & &+\,\left.2\,{^{(2)}\triangle}\,\dot  A - {^{
   (2)}\triangle}\,\dot w \right)+\,4\,{{c_{E}}^2}\,\left( {\cal
   H}\, A - \frac{\kappa}{2}\, \varphi_0' \,\delta\varphi\right) \,w'{} +
   {\frac{{{c_{E}}^2}\,{{w'{}}^2}}{2}} + w\,\left(\left(
   {^{(2)}\triangle} + 4\,{{c_{E}}^2} \right)A \right.\nonumber\\ 
   & & \left.+ 2\,{h_{E}}\,\dot  A \right) + 2\,\dot A\,\dot w -
   \,{\frac{{^{ {(2)}}\triangle}\left(
   \left( 2 + {^{(2)}\triangle} \right) \,{{c_{E}}^2}\,
   {v'{}^2} \right) }{2}}  +
   {}{^{ {(2)}}\triangle}v\, \left(  A\,\left( {^{(2)}\triangle} + 
       4\,{{c_{E}}^2} \right) + 2\,{h_{E}}\,\dot  A \right) \nonumber\\ 
   & &+ \,2\,
   {h_{E}}{}{^{ {(2)}}\triangle}\, A\,\dot v +
     \frac{{^{(2)}\triangle}(2+{^{(2)}\triangle})}{2} \, \dot v^2 +
   2\,{{c_{E}}^2}\,\left( 1 + 2\,{{{\cal H}}^2} + {\cal H}' \right)\,{{ A}^2}
   - 4\,\kappa\,{{a_{E}}^2}\,\delta\varphi\, A\,{{c_{E}}^2}\, 
   V_{,\varphi}\nonumber\\
   & & \left. -\,
   2\,\kappa\,\delta\varphi\,{{c_{E}}^2}\,\varphi_0'\,  A'{} +\,
     \kappa\,{{\delta\varphi}^2}\, \left( {^{
   (2)}\triangle} - {{a_{E}}^2}\,{{c_{E}}^2}\, V_{,\varphi\varphi} \right) -
   \kappa\,{{c_{E}}^2}\,{{\delta\varphi'{}}^2} + \kappa\,\delta\dot
   \varphi^2\right\}.\nonumber
\end{eqnarray}
By a overdot we denote a derivative with respect to $\rho$, and by a
prime we denote a derivative with respect to ${\eta_{E}}$.

The conjugate momenta are introduced as 
\begin{equation}
\label{momout}
 \begin{array}{lcl}
   \Pi_A &=& \displaystyle \frac{a_{E}^2c_{E}\sqrt{\omega}}{\kappa}\left(\dot
     w + h_{E} (w + {{}^{(2)}\triangle}v) + 2 {\cal H}S +
     {^{(2)}\triangle} \xi + 2 h_{E}
     \zeta\right), \\
   \Pi_T &=& \displaystyle - \frac{a_{E}^2 \sqrt{\omega}{^{(2)}\triangle}}{2
     \kappa c_{E}}\left(\dot T - 2 h_{E} T + S + c_{E}^2 \xi'\right),
   \\
   \Pi_w &=& \displaystyle - \frac{a_{E}^2c_{E}\sqrt{\omega}}{2
     \kappa}\left(\dot w - 2 \dot A + 2 S' + 4 {\cal H} S + 2 h_{E}\zeta +
     {^{
         (2)}\triangle} \xi\right), \\
   \Pi_v &=& \displaystyle \frac{a_{E}^2c_{E}\sqrt{\omega}{^{(2)}\triangle}}
   {2\kappa}\left(({^{(2)}\triangle}+2) \dot v + 2 h_{E} A - 
     ({^{(2)}\triangle}
     + 2)\xi\right), \\
   \Pi_{\delta\varphi} &=& \displaystyle
   a_{E}^2c_{E}\sqrt{\omega}\left(\delta\dot\varphi - \varphi_0' S\right).
\end{array}
\end{equation}
The corresponding first order Lagrangian reads:
\begin{eqnarray}
  \label{eq:lagscalout} 
^{(s)}{\cal L} &=& ^{(s)}{\cal L}_1 - ^{(s)}{\cal H} -
   {\cal C}_S S - {\cal C}_\xi \xi - {\cal C}_\zeta \zeta ,\\
 ^{(s)}{\cal L}_1
   &=& \Pi_A\dot  A + \Pi_T \dot T + \Pi_w\dot w + \Pi_v\dot v +
   \Pi_\varphi \delta\dot\varphi \nonumber,\\
 ^{(s)}{\cal H} &=& -\frac{a_{E}^2 c_{E}\sqrt{\omega}}{2
   \kappa}\left\{-{\frac{{^{
   (2)}\triangle}\,{{T}^2}}{{{c_{E}}^2}}} + T\,\left( -4\,{^{
   (2)}\triangle}\,{\cal H}\, A - {\frac{4\,\kappa\,{h_{E}}\,\Pi_T}
   {{{a_{E}}^2}\,c_{E}\,\sqrt{\omega}}} + 2\,\kappa\,{^{
   (2)}\triangle}\,\delta\varphi\, \varphi_0'\right.\right. \nonumber\\
 & & \left.- {^{(2)}\triangle}\,w'{}- {{}^{(2)}\triangle} 
   ( {^{(2)}\triangle} + 2 \frac{}{}) \,v'{} 
   \right) +
   w\,\left( \left( {^{(2)}\triangle} + 4\,{{c_{E}}^2}
   \right) \, A + {\frac{\kappa\,{h_{E}}\,\left( \Pi_A + 2\,\Pi_w \right)
      }{{{a_{E}}^2}\,c_{E}\, \sqrt{\omega}}} \right.
   \nonumber\\
 & & \left.- {{h_{E}}^2}\,{{}^{(2)}\triangle}v\frac{}{}\right)-
   \,{\frac{\left(
   {{h_{E}}}\,{{w}} \right)^2 }{2}}+
   {\frac{4\,\kappa\,{h_{E}}\,\Pi_v\, A} {\left( 2 + {^{
   (2)}\triangle} \right) \,{{a_{E}}^2}\,c_{E}\, \sqrt{\omega}}}+\,4\,
   {{c_{E}}^2}\,
   \left( {\cal H}\, A- \frac{\kappa}{2}\,\varphi_0'\,
   \delta\varphi
   \right) \,w'{} \nonumber\\
 & &+ {}{^{(2)}\triangle}v\,\left( \left( {^{(2)}\triangle} +
   4\,{{c_{E}}^2} \right)\, A + {\frac{\kappa\,{h_{E}}\, \left( \Pi_A +
   2\,\Pi_w \right) }{{{a_{E}}^2}\,R\,\sqrt{\omega}}} \right)-\,
   \frac{{^{(2)}\triangle}\left( {^{
   (2)}\triangle} + 2\right) \,{c_{E}}^2\,
   v'{}^2}{2} \nonumber\\
 & & - {\frac{{h_{E}}^2{{}^{(2)}\triangle}^2\,v^2}{2}}
   + {\frac{2\,{{ A}^2}\,\left( {^{(2)}\triangle} +
   {{c_{E}}^2}\,\left( 2 + 2\,\left( 2 + {^{(2)}\triangle}
   \right) \, {{{\cal H}}^2} + 2\,{\cal H}' + {^{
   (2)}\triangle}\,{\cal H}' \right) \right) } {2 + {^{
   (2)}\triangle}}}
   \nonumber\\
 & & + {\frac{{{c_{E}}^2}\,{{w'{}}^2}}{2}}
   +\kappa\,{{\delta\varphi}^2}\, \left( {^{
   (2)}\triangle} - {{a_{E}}^2}\,{{c_{E}}^2}\, V,_{\varphi\varphi} \right) 
    - 2\,\kappa\,\delta\varphi\,{{c_{E}}^2}\,\varphi_0'\,  A'{}
   - 4\,\kappa\,{{a_{E}}^2}\,
   \delta\varphi\, A\,{{c_{E}}^2}\, V,_{\varphi}\nonumber \\ 
 & & \left.-
   \kappa\,{{c_{E}}^2}\,{{\delta\varphi'{}}^2}\frac{}{}\frac{}{}\right\}-
   \,\frac{\kappa}{a_{E}^2c_{E}\sqrt{\omega}}
   \left\{{\frac{{{{c_{E}}^2}\,{\Pi_T}^2}}{{^{
   (2)}\triangle}}} -  \Pi_A\,\Pi_w -
   {\frac{{{\Pi_v}^2}} {{^{(2)}\triangle}\,(2 +
   {^{(2)}\triangle})}} - {\frac{{{\Pi_A}^2}}{4}} -
   {\frac{{{\Pi_{\delta\varphi}}^2}}{2\,\kappa}}\right\} \nonumber,
\end{eqnarray} 
where
\begin{eqnarray}
  {\cal C}_S &=&  \varphi_0' \Pi_{\delta\varphi}- 2 {\cal H} \Pi_w - \Pi_T 
  - a_{E}\left(\frac{\Pi_A}{a_{E}}\right)' + \frac{a_{E}^2h_{E}c_{E}
    \sqrt{\omega}}{
    \kappa}\left({\cal H} (w + {{}^{(2)}\triangle}v) + {{}^{
        (2)}\triangle}v' + \frac{{^{
          (2)}\triangle}}{c_{E}^2} T\right) \nonumber,\\
  {\cal C}_\xi &=& \Pi_v - {^{(2)}\triangle}\,\Pi_w + c_{E}^2 \Pi_T'
  \nonumber,\\
  {\cal C}_\zeta &=& - 2 {h_{E}}\,\Pi_w - \frac{a_{E}^2 c_{E} \sqrt{\omega}}{
    \kappa}\left(\frac{1}{2}({^{(2)}\triangle} + 2)(w + {{}^{
        (2)}\triangle}v) + 2 c_{E}^2 {\cal H} w' + c_{E}^2 w'' + 2 {\cal
      H}c_{E}^2\,A' - {^{(2)}\triangle} A \nonumber
  \right.\\
  & &+\, \left. 2\,c_{E}^2(2 {\cal H}^2 + {\cal H}' - 1) A- 2 {\cal H} {^{
        (2)}\triangle} T - {^{(2)}\triangle} T' - \kappa a_{E}^2 c_{E}^2
    V_{,\varphi_0} \delta\varphi - \kappa c_{E}^2 \varphi_0'
    \delta\varphi'\frac{}{}\right)\nonumber.
\end{eqnarray}
Notice that when $\ell = 0,1$, equation (\ref{eq:lagscalout}) is meaningless
because of the factors involving the laplacian. For the moment, we assume that
$\ell \neq 0,1$, and postpone the discussion of the fate of this two modes
until the end of this section. 

We note that the fields $\xi$, $\zeta$ and $S$
do not appear in the canonical form ${\cal L}_1$, and only appear linearly in
the lagrangian, i.e., they are $\delta{\cal N}^{\mu}$-like variables. The
constraints are ${\cal C}_S=0$, ${\cal C}_\xi=0$ and ${\cal C}_\zeta=0$, which
generate even parity gauge transformations (see Appendix \ref{sec:gauoutside}).

To reduce the phase space we proceed following the way discussed 
in section \ref{sec:redmethod}. By taking linear combinations of 
the constraints ${\cal C}_S=0$, ${\cal C}_\xi=0$ and 
${\cal C}_\zeta=0$ 
we can construct $\tilde {\cal C}_{\mu}$ which 
takes the form $\tilde {\cal C}_{\mu}=p_{\mu}-\hat p_{\mu}[q]$, 
where $p_{\mu}=\{\Pi_{\delta\varphi}, \Pi_v, \Pi_w\}$ and 
$\{q\}=\{A,\delta\varphi,T,v,w,\Pi_A,\Pi_T\}$,. 
Now we can apply the formula (\ref{giv}) to obtain the 
gauge invariant combinations of variables as 
\begin{equation}
\label{C3}
  \begin{array}{lcl}
    {\bf \Phi} &=&\displaystyle A -
    \frac{1}{a_{E}}\left(a_{E}\,\frac{\delta\varphi}{\varphi_0'}
    \right)' ,\\
    {\bbox{T}} &=&\displaystyle T + \frac{\delta\varphi}{\varphi_0'} + 
    c_{E}^2 v' ,\\
    {\bf \Pi}_{\bf \Phi} &=& \displaystyle\Pi_A - \frac{a_{E}^2
      c_{E}\sqrt{\omega}}{ \kappa h_{E}}\left({\cal H}c_{E}^2{\tilde w}' +
      c_{E}^2 \tilde w + \frac{{^{
            (2)}\triangle}}{2}\tilde w \right) ,\\
    {\bf \Pi}_{\bbox{T}} &=& \displaystyle\Pi_T +
    \frac{a_{E}^2c_{E}\sqrt{\omega}{^{
          (2)}\triangle}}{\kappa h_{E}}\left(\frac{h_{E}^2}{c_{E}^2}\,
      \frac{\delta\varphi}{\varphi_0'} + \frac{\tilde w'}{2}\right),
\end{array}
\end{equation}
where
\begin{equation}
  \tilde w =w+ 2{\cal H}\frac{\delta\varphi}{\varphi_0'} + {{}^{
    (2)}\triangle}v.
\end{equation}

Substituting the constraints into the original first order Lagrangian to
eliminate $p_{\mu}=\{\Pi_{\delta\varphi}, \Pi_v, \Pi_w\}$, we obtain the
Lagrangian that depends only on $q$.  Then, simply taking
$\delta\varphi=v=w=0$ and replacing $A$, $T$, $\Pi_{A}$ and $\Pi_{T}$ by
${\bf \Phi}$, ${\bbox{T}}$, ${\bf \Pi}_{\bf \Phi}$ and 
${\bf \Pi}_{\bbox{T}}$, respectively, we finally get an action which
depends only on the two pair of canonically conjugate gauge invariant
fields:
\begin{eqnarray}
  \label{lagrangianinv}
  ^{(s)}{\cal L}^* &=& {\bf \Pi}_{\bf \Phi}\,\dot{\bf \Phi} 
 + {\bf \Pi}_{\bbox{T}}\,\dot {\bbox{T}} -
  ^{(s)}{\cal H}^*,\\
  ^{(s)}{\cal H}^* &=& {{{\Pi_{\delta\varphi}^{*2}}}\over {2\,{{a_{E}}^2}\,
      c_{E}\,\sqrt{\omega}}}- {{2\,h_{E}\,\Pi_v^*\,{\bf \Phi}
      }\over {{^{(2)}\triangle}({^{(2)}\triangle} + 2)}} + {{\kappa
        \,{\Pi_v^*}^2}\over {{^{(2)}\triangle}\left( {^{
              (2)}\triangle} + 2 \right) \,{{a_{E}}^2}\,c_{E}\,\sqrt{\omega}}}
 + {{\kappa \,\Pi_w^*\,{\bf \Pi}_{\bf \Phi}}\over
      {{{a_{E}}^2}\,c_{E}\,\sqrt{\omega}}}
    \nonumber\\
    & &-\, {{\kappa \,c_{E}\,{{{\bf \Pi}_{\bbox{T}}}^2}}\over {{^{
            (2)}\triangle}\,{{a_{E}}^2}\,\sqrt{\omega}}} - {{2\left( -\kappa
          \,h_{E}\,{\bf \Pi}_{\bbox{T}} - {^{(2)}\triangle}\,{{a_{E}}^2}\,{\cal
           H} \,{{c_{E}}}\,\sqrt{\omega}\,{\bf \Phi} \right) \,{\bbox{T}}}\over
      {\kappa}} + {{{{a_{E}}^2}\,\sqrt{\omega}\,{\bbox{T}}\,{^{
            (2)}\triangle}\,{{\bbox{T}}}}\over {2\,\kappa \,c_{E}}} \nonumber\\
    & &-\, {{{{a_{E}}^2}\,c_{E}\,\sqrt{\omega}\,{{{\bf \Phi} }} \left( {^{
              (2)}\triangle} + {{c_{E}}^2}(2 + ({^{(2)}\triangle} +
          2)(2\,{{{\cal H}}^2} + {\cal H}' ))\right){{\bf \Phi} } }\over
      {\kappa\,\left( {^{(2)}\triangle} + 2 \right) }} 
    +{{\kappa \,{{{\bf \Pi}_{\bf \Phi}}^2}}\over
    {4\,{{a_{E}}^2}\,c_{E}\,\sqrt{\omega}}}  ,\nonumber
\end{eqnarray}
where $\Pi_{\delta\varphi}^*$, $\Pi_v^*$, and $\Pi_w^*$ are functions of
${\bbox{T}}$, ${\bf \Pi}_{\bbox{T}}$, ${\bf \Phi}$ and ${\bf \Pi}_{\bf \Phi}$
determined from the constraints and given by
\begin{equation}
  \begin{array}{lcl}
    \Pi_{\delta\varphi}^* &=& \displaystyle \frac{{\bf \Pi}_{\bbox{T}} +
      {\bf \Pi}'_{\bf \Phi} - {\cal H} \,{\bf \Pi}_{\bf \Phi} }{\varphi_0'} -
    \frac{a_{E}^2 h_{E} \sqrt{\omega}{^{(2)}\triangle} \,{\bbox{T}}}{\kappa
      c_{E}\varphi_0'} + \frac{a_{E}^2{\cal H}c_{E}\sqrt{\omega}}{\kappa h_{E}
      \varphi_0'}{\cal F} ,\\
    \Pi_v^* & = & \displaystyle - c_{E}^2\,{\bf \Pi}_{\bbox{T}}' +
    \frac{a_{E}^2c_{E}{^{(2)}\triangle}\sqrt{\omega}}{2\kappa h_{E}}\,
    {\cal F} ,\\
    \Pi_w^* & =& \displaystyle
    \frac{a_{E}^2c_{E}\sqrt{\omega}}{2\kappa h_{E}}\,{\cal F} ,\\
    {\cal F} &=& \displaystyle {^{(2)}\triangle} {\bf \Phi} + 2\, c_{E}^2 (1
    - 2 {\cal H}^2 - {\cal H}')\,{\bf \Phi} - 2 {\cal H} c_{E}^2\,{\bf
      \Phi}' + {^{(2)}\triangle} (2 {\cal H} \,{\bbox{T}} + {\bbox{T}}').
  \end{array}
\end{equation} 

To disentangle the two degrees of freedom, we proceed in the following way.
This Hamiltonian carries the counterparts of the scalar and even parity tensor
modes inside the lightcone. Since they are decoupled there, we can expect that
they are also decoupled outside the lightcone.  Hence we choose new
coordinates $\{{\bbox{s}}$, ${\bbox{v}}\}$ such that when they are
analytically continued inside the lightcone, they reduce to pure scalar and
pure tensor variables in a particular gauge.  A convenient choice is the
longitudinal gauge for the scalar modes and the synchronous gauge for the
tensor modes, in which $S_i=0$ and the traceless part of the spatial metric
perturbation inside the lightcone becomes a purely tensor quantity. Let us
call this gauge the Newton gauge for convenience.  Then it is easy to see that
${\bf \Phi}$ is a purely scalar type quantity.  For tensor modes, we use the
fact that $S_i=0$ hence $T=0$ and $v$ is a purely tensor type variable in the
Newton gauge.  Then with the help of the equations of motion, we can show that
the following pair of fields ${\bbox{s}}$, ${\bbox{v}}$ become purely scalar
and purely tensor type variables (see Appendix \ref{purely} for a brief
discussion about this subject):
\begin{equation}
  \label{eq:chvar}
  \begin{array}{lcl}
    {\bbox{s}} &=& a_{E} \varphi_0'\,{\bf \Phi}, \nonumber\\
    {\bbox{v}} &=& \displaystyle {\widehat K}[a_{E}{\bbox{T}}] -
    \frac{d}{d{\eta_{E}}}
\left(a_{E}{\bf \Phi}\right).
  \end{array}
\end{equation}
Then we can find new canonically conjugate momenta ${\bf \Pi}_{{\bbox{s}}}$,
${\bf \Pi}_{{\bbox{v}}}$ such that the Hamiltonian $^{(s)}{\cal H}^*$
decouples into two pieces. Just for reference, we recall the definition of the
operators ${\widehat{\cal O}}$ and ${\widehat K}$,
\begin{equation}
  {\widehat{\cal O}} = - \frac{d^2}{d\eta_{E}^2} + \frac{\kappa}{2}
  \varphi_0'{}^2 + \varphi_0'\left(\frac{1}{\varphi_0'}\right)'', \hskip 2cm
  {\widehat K} = - \frac{d^2}{d\eta_{E}^2} + \frac{\kappa}{2} \varphi_0'{}^2.
\end{equation}
It is useful to keep in mind the following relation between ${\widehat{\cal
    O}}$ and ${\widehat K}$:
\begin{equation}
  \label{Ostrel}
  \frac{d}{d\eta_{E}}\frac{1}{\varphi_0'}{\widehat{\cal O}}={\widehat
    K}\frac{1}{\varphi_0'{}^2} \frac{d}{d\eta_{E}}\varphi_0'\,.
\end{equation}

To find the appropriate momenta basis in which the Hamiltonian
decouples, we propose an ansatz for it, namely
\begin{equation}
  \label{eq:chmom}
  \begin{array}{lcl}  
  {\bf \Pi}_{\bf \Phi} &=& a_{E} \varphi_0'\,{\bf \Pi}_{{\bbox{s}}} +
  a_{E}{\bf \Pi}_{\bbox{v}}' + \widehat {\cal
    Q}[{\bf \Phi}] + \widehat {\cal X}_{\bbox{T}}[{\bbox{T}}],\\
  {\bf \Pi}_{\bbox{T}} &=& a_{E}\,{\widehat K}[{\bf \Pi}_{\bbox{v}}] 
+ \widehat {\cal T}[{\bbox{T}}]+ \widehat{\cal X}_{{\bf \Phi}}[{\bf \Phi}],
\end{array}
\end{equation}
where $\widehat {\cal Q}$ and $\widehat {\cal T}$ are in principle arbitrary
differential operators, but $\widehat {\cal X}_{{\bf \Phi}}$ and $\widehat
{\cal X}_{\bbox{T}}$ must be related in order to keep the transformation
canonical.  The momenta dependence of the transformation is found by requiring
the transformation (\ref{eq:chvar})-(\ref{eq:chmom}) to be canonical. Now we
compute the canonical equations of motion for ${\bbox{s}}$ and ${\bbox{v}}$
using the old basis, and express the result in terms of the new basis.  We
find that $\dot{{\bbox{s}}}$ is independent of ${\bf \Pi}_{\bbox{v}}$, and
that $\dot{\bbox{v}}$ is independent of ${\bf \Pi}_{{\bbox{s}}}$. Then we
define the operators involved in the definition of the new momenta basis in
order to completely decouple these two equations.  It can be shown that if we
choose the operators as
\begin{equation}
\label{opa}
  \begin{array}{lcl}
    \widehat{\cal Q}[{\bf \Phi}] &=& \displaystyle -
    \frac{a_{E}^2\,c_{E}\sqrt{\omega}({^{(2)}\triangle}-2({\cal
        H}^2-1)c_{E}^2)}{\kappa h_{E}}\,{\bf \Phi},\\
    \widehat{\cal X}_{\bbox{T}}[{\bbox{T}}] &=& \displaystyle
    -\frac{a_{E}c_{E}\sqrt{\omega}({^{(2)}\triangle} +
      2\,c_{E}^2)}{\kappa h_{E}}\frac{d}{d{\eta_{E}}}a_{E}\,{\bbox{T}}
-\frac{a_{E}^2{\cal H}c_{E}\sqrt{\omega}{^{
          (2)}\triangle}}{\kappa h_{E}}\,{\bbox{T}},
  \end{array}
\end{equation}
\begin{equation}
\label{opb}
  \begin{array}{lcl}
    \widehat{\cal T}[{\bbox{T}}] &=& -
    \displaystyle\frac{a_{E}c_{E}\sqrt{\omega}({^{ (2)}\triangle} +
      2\,c_{E}^2)}{\kappa h_{E}}\,{\widehat K}[a_{E}\,{\bbox{T}} ] -
    \frac{a_{E}^2\sqrt{\omega}{^{(2)}\triangle}}{\kappa
      h_{E}c_{E}}\,{\bbox{T}} ,\\
    \widehat{\cal X}_{{\bf \Phi}}[{\bf \Phi}] &=&\displaystyle
    \frac{a_{E}^2\,c_{E}\sqrt{\omega}({^{(2)}\triangle}+2\,c_{E}^2)}{\kappa
      h_{E}} \frac{d}{d{\eta_{E}}}{\bf \Phi} + \frac{2\,a_{E}^2{\cal
        H}c_{E}^3\sqrt{\omega}}{\kappa h_{E}}{\bf \Phi},
  \end{array}
\end{equation}
$\dot{{\bbox{s}}}$ and $\dot{\bbox{v}}$ turn out to be
\begin{equation}
\label{eq:eqcoord}
\begin{array}{lcl}
  \dot{{\bbox{s}}} &=& \displaystyle \frac{1}{c_{E}\sqrt{\omega}}
  {\widehat{\cal O}}[{\bf \Pi}_{{\bbox{s}}}],\\
  \dot{\bbox{v}} &=& \displaystyle
  \frac{\kappa}{2\,c_{E}\sqrt{\omega}}\left(1-\frac{4\,{\widehat K}c_{E}^2(
      {^{ (2)}\triangle}+2-(1+{\widehat K})c_{E}^2)}{{^{(2)}\triangle}({^{
          (2)}\triangle} + 2)}\right)
  \,{\widehat K}[{\bf \Pi}_{\bbox{v}}] \nonumber\\
  & &\displaystyle +\, \frac{c_{E}^2({^{(2)}\triangle}({^{(2)}\triangle} + 2)
    + 2 (4 + {^{(2)}\triangle}(1 - {\widehat K})) c_{E}^2 - 4 (1+{\widehat
      K})c_{E}^4)}{{^{ (2)}\triangle}({^{(2)}\triangle} + 2)h_{E}} \,{\widehat
    K}[{\bbox{v}}],
\end{array}
\end{equation}
which are already decoupled. It can be verified that the ${\bf
  \Pi}_{{\bbox{s}}}$ and ${\bf \Pi}_{\bbox{v}}$ defined by (\ref{eq:chmom})
with the help of (\ref{opa})-(\ref{opb}) are canonical conjugates of
${\bbox{s}}$ and ${\bbox{v}}$, so the two equations we have computed are two
canonical equation of motion of the system. Therefore, to find the Hamiltonian
in the new basis, we only need to know the two remaining canonical equations
of motion. Computing $\dot{\bf \Pi}_{{\bbox{s}}}$ and $\dot{\bf
  \Pi}_{\bbox{v}}$ we obtain:
\begin{equation}
\label{eq:eqmom}
\begin{array}{lcl}
  {\widehat{\cal O}}[\dot {\bf \Pi}_{{\bbox{s}}}] &=& \displaystyle
  c_{E}^3\sqrt{\omega}\,\left[(3 +
    \frac{{^{(2)}\triangle}}{c_{E}^2})\,{\bbox{s}} - {\widehat{\cal O}}[
    {\bbox{s}}]\right]\\
  {\widehat K}[\dot {\bf \Pi}_{\bbox{v}}] &=& \displaystyle
  \frac{2\,c_{E}^5\sqrt{\omega}{\widehat K}}{\kappa}\,\left(\frac{{^{
          (2)}\triangle}(4+{^{(2)}\triangle}(1-{\widehat K})) - 4
      c_{E}^2({\widehat K} {^{(2)}\triangle}-2 + (1+{\widehat
        K})c_{E}^2)}{{^{(2)}\triangle}({^{
          (2)}\triangle}+2)h_{E}^2}\right.  \nonumber\\
  & & \displaystyle\left. +\frac{ {^{(2)}\triangle}+3c_{E}^2}{c_{E}^4{\widehat
        K}}\right)\,{\bbox{v}} - \frac{{\widehat K}^2c_{E}^2}{{^{
        (2)}\triangle}({^{(2)}\triangle} + 2)h_{E}}\left(2 (4 +
    {^{(2)}\triangle}(1 - {\widehat K})) c_{E}^2 \right.\\
  & & \displaystyle \left.+{^{(2)}\triangle}({^{(2)}\triangle} + 2)-
    4(1+{\widehat K})c_{E}^4\right) \,{\bf \Pi}_{\bbox{v}}
\end{array}
\end{equation}
Note that all the ${\eta_{E}}$ dependence has been absorbed in the
differential operators ${\widehat{\cal O}}$ and ${\widehat K}$. Expanding
(\ref{eq:eqcoord}) and (\ref{eq:eqmom}) in terms of eigenfunctions of this
operators, we can read directly from them the coefficients of the Hamiltonian
in the new basis. The corresponding first order lagrangian is
\begin{eqnarray}
  ^{(s)}{\cal L}^* &=& {\cal L}_{{\bbox{q}}}[{{\bbox{s}}},{\bf
    \Pi}_{{\bbox{s}}}] + {\cal L}_{\bbox{w}}
  [{\bbox{v}},{\bf \Pi}_{\bbox{v}}]\label{eq:lagout},\\
  {\cal L}_{{\bbox{q}}} &=& {\bf \Pi}_{{\bbox{s}}} \dot{{{\bbox{s}}}} - {\bf
    \Pi}_{{\bbox{s}}}\frac{{\widehat{\cal O}}}{2 c_{E} \sqrt{\omega}}{\bf
    \Pi}_{{\bbox{s}}} +
  \frac{c_{E}^3\sqrt{\omega}}{2}\,{\bbox{s}}\,\frac{1}{{\widehat{\cal
        O}}}\left(3 + \frac{{^{ (2)}\triangle}}{c_{E}^2} -
    {\widehat{\cal O}}\right){\bbox{s}}\nonumber,\\
  {\cal L}_{{\bbox{w}}} &=& {\bf \Pi}_{\bbox{v}} \dot {\bbox{v}} - \frac{1}{2}
  {\bf \Pi}_{\bbox{v}}\,\frac{\kappa{\widehat
      K}}{2\,c_{E}\sqrt{\omega}}\left(1-\frac{4\, {\widehat
        K}c_{E}^2({^{(2)}\triangle}+2-(1+{\widehat
        K})c_{E}^2)}{{^{(2)}\triangle}({^{ (2)}\triangle} + 2)}\right) \,{\bf
    \Pi}_{\bbox{v}}
  \nonumber\\
  & &-\,{\bf \Pi}_{\bbox{v}} \,{\widehat
    K}\frac{c_{E}^2({^{(2)}\triangle}({^{(2)}\triangle} + 2) + 2 (4 +
    {^{(2)}\triangle}(1 - {\widehat K})) c_{E}^2 - 4 (1+{\widehat
      K})c_{E}^4)}{{^{ (2)}\triangle}({^{
        (2)}\triangle} + 2)h_{E}} \,{\bbox{v}} \nonumber\\
  & &+\,
  \frac{1}{2}\,{\bbox{v}}\,\frac{2\,c_{E}^5\sqrt{\omega}}{\kappa}\,\left
    (\frac{{^{(2)}\triangle}(4+{^{ (2)}\triangle}(1-{\widehat K})) - 4c_{E}^2
      ({^{(2)}\triangle}{\widehat K}-2 + (1+{\widehat
        K})c_{E}^2)}{{^{(2)}\triangle}({^{
          (2)}\triangle}+2)h_{E}^2}\right.\nonumber\\
  & &\left.+\frac{{^{ (2)}\triangle}+3c_{E}^2}{{\widehat
        K}c_{E}^4}\right)\,{\bbox{v}}.\nonumber
\end{eqnarray}

The lagrangian ${\cal L}_{{\bbox{s}}}$ can be put easily in second order form.
Solving for the momenta ${\bf \Pi}_{{\bbox{s}}}$, and defining ${\bbox{q}}$
through
\begin{equation}
  \label{C16}
  {\bbox{s}} = {\widehat{\cal O}}[{\bbox{q}}],
\end{equation}
we find equation (\ref{eq:mainoutsc}) of the text.

The lagrangian ${\cal L}_{{\bbox{w}}}$ needs a little extra work. Performing
the following canonical transformation
\begin{equation}
\label{lcl}
\begin{array}{lcl}
  {\bbox{w}} &=& \displaystyle \frac{h_{E}}{c_{E}\sqrt{\omega}}{\bf
    \Pi}_{\bbox{v}} -
  \frac{{^{(2)}\triangle}+2\,c_{E}^2}{\kappa{\widehat K}}{\bbox{v}},\\
  {{\bf \Pi}_{\bbox{w}}} &=& \displaystyle \frac{\kappa{\widehat K}(2
    ({\widehat K}c_{E}^2 - h_{E}^2) -{^{(2)}\triangle})}{{^{(2)}\triangle}({^{
        (2)}\triangle}+2)}{\bf \Pi}_{\bbox{v}} +
  2c_{E}\sqrt{\omega}\,\frac{2h_{E}^2({^{(2)}\triangle} + c_{E}^2) - {\widehat
      K}\,({^{ (2)}\triangle} + 2\,c_{E}^2)c_{E}^2}{ {^{(2)}\triangle}({^{
        (2)}\triangle}+2)h_{E}}{\bbox{v}},
\end{array}
\end{equation}
we find for ${\cal L}_{\bbox{w}}$
\begin{equation}
  {\cal L}_{\bbox{w}} = {\bf \Pi}_{\bbox{w}}\,\dot{\bbox{w}} +
  \kappa{\bbox{w}}{\widehat K}\ \frac{c_{E}\sqrt\omega({^{(2)}\triangle} -
    (1+{\widehat K})c_{E}^2 )}{{^{(2)}\triangle}({^{
        (2)}\triangle}+2)}{\bbox{w}} - {\bf \Pi}_{\bbox{w}}\frac{{^{
        (2)}\triangle}({^{ (2)}\triangle}+2)}{4\kappa{\widehat
      K}c_{E}\sqrt{\omega}}{\bf \Pi}_{\bbox{w}}\,.
\label{Lagw}
\end{equation}
Solving for the momenta ${\bf \Pi}_{\bbox{w}}$, we find equation
(\ref{eq:mainoutvec}) in the text.

When $\ell=0,1$, simply looking at the definition of momenta (\ref{momout}),
we can see that new constraints arises, due to the fact that some of them
vanish.  For $\ell=0$, $\Pi_T$ and $\Pi_v$ become zero. In fact, the second
order lagrangian $^{(s)}{\cal L}$ for $\ell=0$ is independent of $\xi$, $T$
and $v$. In this case we are left with a Lagrangian that depends only on three
fields plus two lagrangian multipliers, therefore $^{(s)}{\cal L}_{\ell=0}$
only contains one degree of freedom.  Applying the Faddev-Jackiw formalism,
the action for this degree of freedom turns out to be the one for the scalar
degree of freedom, ${\cal S}^{(2)}_{\bbox{q}}$, with $\ell=0$. Similarly, if
$\ell=1$ the lagrangian $^{(s)}{\cal L}_{\ell=1}$ is independent of $v$, so we
have four fields and three lagrangian multipliers.  As before, $^{(s)}{\cal
  L}_{\ell=1}$ only contains one degree of freedom. As expected, in this case
we recover the action ${\cal S}^{(2)}_{\bbox{q}}$ for $\ell=1$. This is
consistent with the fact that ${\bbox{w}}$ represents one of the tensor
degrees of freedom inside the lightcone, so it must be absent for $\ell=0,1$.
On the other hand, ${\bbox{q}}$ represents the scalar degree of freedom, so it
must exists for all $\ell$.

\section{Guessing the new variables}
\label{purely}
As we have said, the guess for the variables 
which disentangle the lagrangian
(\ref{lagrangianinv}) is motivated by their expression when 
evolved inside the lightcone in a particular gauge 
(Newton gauge, defined in Appendix
\ref{sec:FJscalar}), where we know they are purely scalar-type or 
purely tensor-type variables. To show this, we need to derive 
some useful and well known
relations between the scalar potentials\footnote{The potential 
${\bf \Phi}_{\bf A}$ is given by ${\bf \Phi}_{\bf A}:= A 
+ (a(B-E'))'/a$, as defined by Bardeen \cite{pertth}.}
 ${\bf \Phi}_{\bf H}$ and ${\bf \Phi}_{\bf A}$. 
By a subscript (or superscript) $N$ we
indicate that the quantity is evaluated in the Newton gauge.

{}First we consider the variables inside the lightcone. 
In the Newton gauge $B_N$ and $E_N$ vanish, so the constraint ${\cal C}_B=0$
reduces to
\begin{equation}
  \label{cbconst}
  \psi_N' - \frac{\kappa}{2}\varphi_0'\delta\varphi_N + {\cal H}A_N = 0.
\end{equation}
Substituting it into the definition of $\Pi_\psi$, Eq. (\ref{defppsi}),
we find that this momentum also vanishes,
\begin{equation}
  \label{ppsin}
  \Pi_{\psi}^N = 0.
\end{equation}
Expressing the equations of motion for ${\bf \Pi}_{\bf \Psi}$ 
and ${\bf \Psi}$, which follow from Eq.~(\ref{B12}), 
in terms of $\psi_N$ and $\delta\varphi_N$, and
eliminating $\delta\varphi_N'$ between them, we find
\begin{equation}
  \label{const}
  \frac{1}{a}(a\,\psi_N)' - \frac{\kappa}{2}\varphi_0'\delta\varphi_N = 0.
\end{equation}
Comparing with (\ref{cbconst}), we recover the well known relation
\begin{equation}
  \label{relatpot}
  ({\bf \Phi}_{\bf A}=) A_N = \psi_N (= - {\bf \Phi}_{\bf H}).
\end{equation}
The analytic continuation of these relations (\ref{const}) and 
(\ref{relatpot}) to the outside of the lightcone does not change their form. 

Now, returning to the outside of the lightcone, 
we justify our choice of variables. In the Newton gauge, 
evaluating ${\bf \Phi}$ defined in 
(\ref{C3}) with the aid of (\ref{const}) and (\ref{relatpot}), we find
\begin{eqnarray}
  \label{eq:phiin}
  {\bf \Phi} = \frac{1}{a_E\varphi_0'}{\widehat{\cal
      O}}\left[\frac{2a_E}{\kappa\varphi_0'} \psi_N \right]
\end{eqnarray}
so ${\bf \Phi}$, and therefore ${{\bbox{s}}}$ and ${\bbox{q}}$, are already a 
purely scalar-type variable. 
Recalling the definition of $\bbox{q}$ given by Eqs.~(\ref{eq:chvar}) and 
(\ref{C16}):${\bf \Phi} = {\widehat{\cal O}}[{\bbox{q}}]
/(a_{E}\varphi_0')$, we find 
\begin{equation}
 {\bbox{q}} = {{2a_{E}}\psi_N\over {\kappa\varphi_0'}}.
\label{qtoPhiH}
\end{equation} 
By means of the fact ${\bf \Phi}_{\bf H}=-\psi_N$ inside the lightcone, 
we find Eq.~(\ref{eq:qpsi}) in the text. 

To find the tensor-type variable, we use the fact that in the 
Netwton gauge $v_N$ is a purely tensor-type variable. 
The strategy is to construct a combination of ${\bf \Phi}$
and ${\bbox{T}}$ proportional to $v_N$. 
First, recalling that $T_N=0$, we evaluate $\bbox{T}$ 
defined in (\ref{C3}) in the Newton gauge as 
\begin{equation}
 \bbox{T}=\frac{2}{\kappa a_E \varphi_0'{}^2}(a_E\psi_N)'+ c_{E}^2 \,v_N'. 
\end{equation}
Then if we define 
\begin{equation}
  \label{defv}
  \tilde{\bbox{T}}:=a_E{\bbox{T}} 
  -\frac{2}{\kappa\varphi_0'{}^2}(a_E\psi_N)' =
  a_Ec_{E}^2\,v_N', 
\end{equation}
we find this quantity becomes a purely tensor-type variable. 
Acting with ${\widehat K}$, we finally find the desired variable
\begin{eqnarray}
  \label{defkv}
  {\bbox{v}} &:=& {\widehat K}[\tilde{\bbox{T}}] = {\widehat K}[a_E{\bbox{T}}]
  - \frac{d}{d\eta}\frac{1}{\varphi_0'}{\widehat{\cal O}}[{2a_E\over
    \kappa\varphi'_0}\psi_N] \cr &=& {\widehat K}[a_E{\bbox{T}}] -
  \frac{d}{d\eta} (a_E{\bf \Phi}),
\end{eqnarray}
where we have used the relation (\ref{Ostrel}). 

\section{Gauge transformations outside the lightcone}
\label{sec:gauoutside}

We can verify that under an even parity gauge transformation generated by $
\lambda^0 \,{\cal C}_S + \lambda^\rho \,{\cal C}_\zeta + \lambda \,{\cal
C}_\xi$, i.e. a diffeomorphism $x^\mu\rightarrow x^\mu+\lambda^\mu$ with
$\lambda^\mu = (\lambda^0,\lambda^\rho,\lambda^{||A})$, the
canonical scalar fields transform according to
\begin{equation}
  \label{eq:transfsc}
 \begin{array}{ll} 
\delta_g  A = \lambda^0{}' +{\cal H} \lambda^0,\quad 
    &\delta_g T= -\lambda^0 - c_{E}^2\lambda',\\
    \delta_g w = - 2 {\cal H} \lambda^0 - 2 h_{E}\lambda^\rho 
     - {^{(2)}\triangle} \lambda,\quad 
 & \delta_g v = \lambda, \\
  \delta_g \delta\varphi = \varphi_0'\,\lambda^0 .&\\
\end{array}
\end{equation} 
The scalar constraints and the scalar Hamiltonian $^{(s)}{\cal H}$
satisfy the following algebra:
\begin{equation}
   \label{salgebra}
  \begin{array}{lll} 
    \{^{(s)}{\cal H},{\cal C}_S\} = \partial_\rho{\cal C}_S
    - {\cal H} {\cal C}_\zeta , & \hskip 1cm &\{^{(s)}{\cal H},{\cal
      C}_\zeta\} = \partial_\rho{\cal C}_\zeta - c_{E}^2 {\cal C}_S' 
  - h {\cal C}_\zeta + {\cal C}_\xi , \\
    \{^{(s)}{\cal H},C_\xi\} = \partial_\rho {\cal C}_\xi, & \hskip 1cm &
    \{{\cal C}_\alpha,{\cal C}_\beta\} = 0. 
  \end{array}
\end{equation}
Notice that the partial derivative with respect to $\rho$ acting on the
constraints only affects background quantities.
Using Eqs.~(\ref{eq:transfsc}) and (\ref{salgebra}), and the fact that
the action (\ref{eq:lagscalout}) is invariant under a gauge
transformation, we find the transformation law for the lagrangian
multipliers as 
\begin{equation}
  \label{eq:transfmultsc} 
 \begin{array}{ccc} \delta_g S = \dot\lambda^0 
    - c_{E}^2\lambda^\rho{}',~ & ~ \delta_g \xi = \dot \lambda -
    \lambda^\rho, ~ & ~ \delta_g \zeta = \dot \lambda^\rho +
    h_{E}\lambda^\rho + {\cal H}\lambda^0. \end{array}
\end{equation}

For the odd parity modes,
under a diffeomorphism generated by $\lambda^\mu=(0,0,\lambda^A)$, where
$\lambda^A$ is divergenceless; $\lambda^A{}_{||A}=0$, 
the fields transform according to
\begin{equation}
\label{eq:transfvec}
  \begin{array}{ccc} \delta_g V_A = - c_{E}^2 \lambda'_A, 
& \hskip 1cm & \delta_g F_A = - \lambda_A \,.
  \end{array}
\end{equation}
The algebra satisfied by the Hamiltonian $^{(v)}{\cal H}$ and the
constraint is:
\begin{equation}
  \{^{(v)}{\cal H},{\cal C}_W^A\}= \partial_\rho {\cal C}_W^A.
\end{equation}
The action is invariant if we transform $W$ as
\begin{equation}
\label{eq:transfmultvec}
  \delta_g W_A = \dot\lambda_A.
\end{equation}


\begin{thebibliography}{10}
  
\bibitem{models} M. Sasaki, T. Tanaka, K. Yamamoto and J. Yokoyama, Phys.
  Lett. B {\bf 317}, 510 (1993); K. Yamamoto, M. Sasaki and T. Tanaka,
  Astrophy. J. {\bf 455}, 412 (1995); M. Bucher, A. Goldhaber and N. Turok,
  Nucl. Phys. {\bf B}, Proc. Suppl.  {\bf 43}, 173 (1995); M. Bucher and N.
  Turok, Phys.  Rev. {\bf D52}, 5538 (1995); A.  Linde. Phys. Lett. B {\bf
    351}, 99 (95); A. Linde and A. Mezhlumian, Phys. Rev. {\bf D52}, 6789
  (1995); A.M. Green and A.R. Liddle, Phys. Rev. {\bf D55}, 609 (1997).

\bibitem{ratra} K.M. Gorski, B. Ratra, R. Stompor and N. Sugiyama and 
A. J. Banday, astro-ph/9608054; 
K. Ganga, B. Ratra, J. O. Gundersen, N. Sugiyama, astro-ph/9602141.

\bibitem{topdef} A. Vilenkin, astro-ph/9703201.

\bibitem{ratra2} B. Ratra and P. Peebles, Phys. Rev. {\bf D52}, 1837
  (1995).

\bibitem{sqfopen} K. Yamamoto, M. Sasaki and T, Tanaka,
  Phys. Rev. {\bf D54}, 5031 (1996). 
  
\bibitem{cananis} M. Sasaki and T. Tanaka. Phys. Rev. {\bf D54}, 4705
  (1996); J. Garriga and V. F. Mukhanov, astro-ph/9702201.
      
\bibitem{beldle} J. Garcia-Bellido and A.R. Liddle, Phys. Rev. 
  {\bf D55}, 4603 (1997); J. Garcia-Bellido and A. Linde,
  Phys. Rev. {\bf D55}, 7480 (1997).

\bibitem{cohn} J.D. Cohn, Phys. Rev. {\bf D54}, 7215 (1996).

\bibitem{allen} B. Allen and R. Caldwell, report No. WISC-MILW-94-TH-21
  (1994).

\bibitem{tamamilne} T. Tanaka and M. Sasaki, Phys. Rev. {\bf D55}, 6061
  (1997).

\bibitem{tamaopen} T. Tanaka and M. Sasaki, Prog. Theor. Phys. {\bf 97},
  243 (1997).

\bibitem{gwbuch} M. Bucher and J.D. Cohn, Phys. Rev. {\bf D55}, 7461
  (1997).

\bibitem{tamaanis} M. Sasaki, T. Tanaka and Y. Yakushige,
astro-ph/9702174, to be published in Phys. Rev. D.

\bibitem{bell1} J. Garcia-Bellido, astro-ph/9702211. 

\bibitem {k3} T. Hamazaki, M. Sasaki, T. Tanaka and K. Yamamoto. Phys. Rev.
  {\bf D53}, 2045 (1996)

\bibitem{bell2} J. Garcia-Bellido, Phys. Rev. {\bf D54}, 2473 (1996).

\bibitem{jaume} J. Garriga, Phys.  Rev. {\bf D54} 4764 (1996).
  
\bibitem{bardeen} J.~M. Bardeen, Phys. Rev. {\bf D22}, 1882 (1980).

\bibitem{kodsas} H. Kodama and M. Sasaki, Prog. Theor, Phys. Suppl. 
{\bf78}, 1 (1984).

\bibitem{pertth} J.~Stewart, Class. Quant. Grav. {\bf 7}, 1169 (1990); 
M.~Bruni, G.~Ellis, and P.~Dunsby, Class. Quant. Grav. {\bf 9}, 921 (1992).

\bibitem{branden}  V.F.~Mukhanov, H.A.~Feldmann and R.H.~Brandenberger,
Phys. Rep. {\bf 115}, 203 (1992).

\bibitem{qflat} M. Sasaki, Prog. Theor. Phys. {\bf 76}, 1036 (1986);
N. Makino and M. Sasaki, Prog. Theor. Phys. {\bf 86}, 103 (1991).

\bibitem{mukhanov} V.F. Mukhanov, Zh. Eksp. Teor. Fiz. {\bf 94}, 1
(1988); S. Anderegg, V.F. Mukhanov, Phys. Lett. B {\bf 331}, 30 (1994).

\bibitem{qclose} T. Tanaka and M. Sasaki, Prog. Theor. Phys. {\bf 88},
503 (1992).

\bibitem{nocauchy} M. Sasaki, T. Tanaka and K. Yamamoto,
Phys. Rev. {\bf D51}, 2979 (1995).

\bibitem{Dirac} P.A.M. Dirac, \newblock {\em Lectures on Quantum Mechanics}.
\newblock Yeshiva University, 1964.
 
\bibitem{tears} L.~Faddeev and R.~Jackiw, Phys. Rev. Lett {\bf60}, 1692
(1988);
R.~Jackiw, in \newblock {\em Diverse Topics in Theoretical and
Mathematical Physics}, \newblock (Constraint) Quantization without
Tears (World Scientific, Singapore, 1995).

\bibitem{pons} J.A. Garc{\'\i}a and J.M. Pons, Int. J. Mod. Phys. {\bf
    A12}, 451 (1997).

\bibitem{tomita} K. Tomita, Prog. Theor. Phys {\bf 68}, 310 (1982).

\bibitem{stspec} J. Garriga, X. Montes, M. Sasaki and T. Tanaka, in
preparation.

\bibitem{ishish} A. Ishibashi and H. Ishihara, gr-qc/9704058.

\bibitem{ADM} Arnowitt, Deser and Misner.  \newblock {\em Gravitation: an
Introduction to Current Research}.  \newblock The Dynamics of General
Relativity. John Wiley.

\bibitem{wipf} For a discussion of symmetries in the first order formalism,
see V.F.~Mukhanov and A. Wipf, Int. J. Mod. Phys. {\bf A10}, 579 (1995).

\end{thebibliography}
\end{document}